\documentclass[aps,prd,twocolumn,nofootinbib,groupedaddress,amsfonts,float	x]{revtex4}

\usepackage{graphicx,amsmath,amssymb,amstext}
\usepackage{amssymb,amsbsy,amsfonts,amsthm,hyperref}
\usepackage{lscape,placeins}
\usepackage{float}

\usepackage{color}
\usepackage{ifthen}
\newboolean{editorial}
\setboolean{editorial}{true}
\newcommand{\editorial}[2]{\ifthenelse{\boolean{editorial}}{\textcolor{red}{[\textsf{\textbf{{#1}}}: }\textcolor{blue}{\textsf{{#2}}}\textcolor{red}{]}}{}}

\begin{document}

\title{Integration of inhomogeneous cosmological spacetimes in the BSSN formalism}

\author{James B. Mertens${}^{1}$}
\author{John T. Giblin, Jr${}^{1,2}$}
\author{Glenn D. Starkman${}^{1}$}

\affiliation{${}^1$CERCA/ISO, Department of Physics, Case Western Reserve University, 10900 Euclid Avenue, Cleveland, OH 44106}
\affiliation{${}^2$Department of Physics, Kenyon College, 201 N College Rd, Gambier, OH 43022}

\begin{abstract}
We present cosmological-scale numerical simulations of an evolving universe in full general relativity (GR) and introduce a new numerical tool, {\sc CosmoGRaPH}, which employs the Baumgarte-Shapiro-Shibata-Nakamura (BSSN) formalism on a 3-dimensional grid.  Using {\sc CosmoGRaPH}, we calculate the effect of an inhomogeneous matter distribution on the evolution of a spacetime.  We also present the results of a set of standard stability tests to demonstrate the robustness of our simulations.
\end{abstract}

\maketitle

\section{Introduction}
\label{intro}

Current cosmological work typically relies on a perturbative approach or a Newtonian-gravity approximation
\cite{Bardeen:1980kt,Kodama:1985bj}. 
Although such approximate methods have led to very precise simulations,
as we acquire and anticipate sub-percent-level precision measurements of the structure in and of the present-day universe, more accurate simulations and therefore more accurate methods are needed. This is not only to provide more accurate results, but also to yield insights into physical processes not previously appreciated.

In this paper we present a numerical implementation of the fully general-relativistic BSSN formalism for studying gravitational systems \cite{Baumgarte:1998te,Shibata:1995we}. This formalism is most often applied in regimes of strong gravity -- compact object dynamics.  However, it has also been applied to cosmological problems in the early universe, such as critical collapse in a radiation fluid \cite{Clough:2015sqa,Baumgarte:2015aza}, black hole lattices (eg. \cite{Bentivegna:2012ei}), and more (for a recent review, see \cite{Cardoso:2014uka}).
Here we examine the performance of the BSSN formalism in the context of a matter-dominated cosmological spacetime, and draw comparisons between FLRW spacetimes and perturbed FLRW spacetimes.

The equations we evolve are fully nonlinear parameterizations of GR, and therefore formidable to work with analytically -- one reason approximations are commonly made. Nevertheless, the nonlinear terms are few enough that, depending on gauge choice, numerically integrating the full unconstrained Einstein equations does not require significantly more computational resources than working in a linearized gravity regime.
In this paper we make the case that nonlinear gravitational effects can be taken into account in cosmological simulations using such a formalism, studying whether a fully unconstrained, general relativistic model of a matter-dominated inhomogeneous universe is consistent with FLRW approximations.

Although the system that we study does not depend on an averaging scheme or background solution, there are several assumptions we make in order to simplify numerical integration. The most significant of these are that we use a pressureless fluid with zero initial coordinate velocity, limiting the dynamics we can probe in the matter sector; the simulation contains a short-wavelength cutoff due to finite available resolution; and the system contains a long-wavelength cutoff due to periodic boundary conditions. Here we examine a numerical spacetime beyond the FLRW approximation in synchronous gauge subject to these constraints, obtaining a full nonlinear solution in a cosmological context. To our knowledge, this is the first time that such a formalism has been applied to an unconstrained, inhomogeneous, 3+1-dimensional cosmological spacetime with a pressureless fluid.

In this work we first describe a simple way to determine initial conditions for a universe containing matter and a cosmological constant term. We then numerically evolve a system described by these initial conditions, using a mildly modified version of the standard BSSN equations. We discuss the subsequent evolution of the system, limitations we encounter, and some potential directions for future work.

\subsection{The BSSN Formalism}

The BSSN equations are derived from the original ADM formalism \cite{Arnowitt:1959ah} of GR, and provide a numerically stable scheme for evolving Einstein's equations. (For a full textbook treatment, see \cite{BaumgarteShapiroBook}.) In this formulation, the metric is written as
\begin{equation}
g_{\mu\nu}=\left(\begin{array}{cc}
-\alpha^{2}+\gamma_{lk}\beta^{l}\beta^{k} & \beta_{i}\\
\beta_{j} & \gamma_{ij} 
\end{array}\right)\,,
\end{equation}
where $\alpha$ and $\beta_{i}$ are referred to as the `lapse' and the `shift' respectively.
The BSSN equations evolve the 3-metric for a particular gauge choice, along with the extrinsic curvature $K_{ij}$.  
The metric is rescaled by a conformal factor,  $\gamma_{ij} \equiv e^{4\phi}\bar{\gamma}_{ij}$, with $\det(\bar{\gamma}_{ij}) = 1$.  
The extrinsic curvature is decomposed into a trace, $K$, and a conformally related trace-free part, $\bar{A}_{ij}$, whose indices are raised and lowered by the conformal metric $\bar{\gamma}_{ij}$;
\begin{equation}
K_{ij} = e^{4\phi}\bar{A}_{ij} + \frac{1}{3}\gamma_{ij}K.
\end{equation}
The full dynamical system is
\begin{eqnarray}
\partial_{t}\phi & = & -\frac{1}{6}\alpha K+\beta^{i}\partial_{i}\phi+\frac{1}{6}\partial_{i}\beta^{i}\\
\nonumber
\partial_{t}\bar{\gamma}_{ij} & = & -2\alpha\bar{A}_{ij}+\beta^{k}\partial_{k}\bar{\gamma}_{ij}\\&&+\bar{\gamma}_{ik}\partial_{j}\beta^{k}+\bar{\gamma}_{kj}\partial_{i}\beta^{k}-\frac{2}{3}\bar{\gamma}_{ij}\partial_{k}\beta^{k}\\
\nonumber
\partial_{t}K & = & -\gamma^{ij}D_{j}D_{i}\alpha+\alpha(\bar{A}_{ij}\bar{A}^{ij}+\frac{1}{3}K^{2})\\
&&+4\pi\alpha(\rho+S)+\beta^{i}\partial_{i}K\\
\nonumber
\partial_{t}\bar{A}_{ij} & = & e^{-4\phi}(-(D_{i}D_{j}\alpha)+\alpha(R_{ij}-8\pi S_{ij}))^{TF}\\
\nonumber
&&+\alpha(K\bar{A}_{ij}-2\bar{A}_{il}\bar{A}_{j}^{l}) +\beta^{k}\partial_{k}\bar{A}_{ij}\\&&+\bar{A}_{ik}\partial_{j}\beta^{k}+\bar{A}_{kj}\partial_{i}\beta^{k}-\frac{2}{3}\bar{A}_{ij}\partial_{k}\beta^{k} \,.
\end{eqnarray}
The source terms written in terms of the stress-energy tensor $T^{\mu\nu}$ are
\begin{eqnarray}
\rho & = & n_\mu n_\nu T^{\mu\nu}\\
S_i & = & -\gamma_{i\mu} n_\nu T^{\mu\nu} \\
S_{ij} & = & \gamma_{i\mu} \gamma_{j\nu} T^{\mu\nu}\,,
\end{eqnarray}
where $n_\mu = (-\alpha, \vec{0})$ and $S = \gamma^{ij}S_{ij}$.

Our software, the Cosmological General Relativistic and Perfect-Fluid Hydrodynamics ({\sc CosmoGRaPH}) code, allows for an arbitrary lapse and shift. However, the work we present here largely makes use of synchronous gauge, or geodesic slicing.
In this gauge, the lapse is a fixed constant ($\alpha = 1$) and there is no shift ($\beta^i = 0$). 
With this choice, the evolution equations for the metric reduce to
\begin{eqnarray}
\partial_{t}\phi & = & -\frac{1}{6} K\\
\partial_{t}\bar{\gamma}_{ij} & = & -2\bar{A}_{ij}\\
\partial_{t}K & = & \bar{A}_{ij}\bar{A}^{ij}+\frac{1}{3}K^{2}+4\pi(\rho+S)+\beta^{i}\partial_{i}K\\
\partial_{t}\bar{A}_{ij} & = & e^{-4\phi}(R_{ij}-8\pi S_{ij})^{TF}+K\bar{A}_{ij}
-2\bar{A}_{il}\bar{A}_{j}^{l}\,.\quad\quad
\end{eqnarray}

For a flat FLRW solution to Einstein's equations, the BSSN variables can be directly translated to the FLRW metric functions.
The FLRW spatial metric is $\gamma_{ij} = a^{2}\delta_{ij}$, meaning $\gamma=\det\gamma_{ij}=a^{6}$.  
Inserting this relationship into the BSSN equations and applying the gauge choice will give us a translation between BSSN and FLRW parameters: $H \sim -\frac{1}{3}K$ and $a \sim \gamma^{1/6} = e^{2\phi}$.
These will be useful for computing initial conditions, and for making comparisons with standard results.

In the BSSN formalism, auxiliary conformal connection variables,
\begin{equation}
\bar{\Gamma}^i\equiv \bar{\gamma}^{jk}\bar{\Gamma}^{i}_{jk}\,,
\end{equation}
with $\bar{\Gamma}_{ijk}=\bar{\gamma}_{il}\bar{\Gamma}_{jk}^{l}$, are evolved simultaneously with the metric functions. These are used to eliminate terms with mixed derivatives when calculating the Ricci tensor.
One writes
\begin{equation}
R_{ij}=\bar{R}_{ij}+R_{ij}^{\phi}
\end{equation}
where 
\begin{eqnarray}
\bar{R}_{ij}&\equiv&
	-\frac{1}{2}\bar{\gamma}^{lm}\partial_{m}\partial_{l}\bar{\gamma}_{ij}
	+\bar{\gamma}_{k(i}\partial_{j)}\bar{\Gamma}^{k}+\bar{\Gamma}^{k}\bar{\Gamma}_{(ij)k}\\
	&&+\bar{\gamma}^{lm}\left(2\bar{\Gamma}_{l(i}^{k}\bar{\Gamma}_{j)km}+\bar{\Gamma}_{im}^{k}\bar{\Gamma}_{klj}\right)\nonumber
\end{eqnarray} 
and 
\begin{equation}
R_{ij}^{\phi}\equiv
	-2\bar{D}_{i}\bar{D}_{j}\phi-2\bar{\gamma}_{ij}\bar{D}^{2}\phi
	+4\bar{D}_{i}\phi\bar{D}_{j}\phi-4\bar{\gamma}_{ij}\left(\bar{D}\phi\right)^{2}\,,
\end{equation} 
where $\bar{D}$ is the covariant derivative associated with $\bar{\gamma}_{ij}$.

The equation of motion for $\bar{\Gamma}^i$ is
\begin{equation}
\partial_{t}\bar{\Gamma}^{i}=2\left(\bar{\Gamma}_{jk}^{i}\bar{A}^{jk}-\frac{2}{3}\bar{\gamma}^{ij}\partial_{j} K-8\pi\bar{\gamma}^{ij}S_{j}+6\bar{A}^{ij}\partial_{j}\phi\right)\,.
\end{equation}
While it might seem odd that we evolve so many (seemingly redundant) parameters, this strategy helps provide numerical stability in the BSSN framework.
By evolving the metric coefficients, extrinsic curvature and connections separately, 
we can check for consistency throughout the simulations. 
These constraints are referred to as the Hamiltonian and momentum constraints, 
and should be obeyed on spacetime slices both initially and throughout the simulation. 
In terms of BSSN variables the Hamiltonian and momentum constraints are respectively
\begin{align}
\nonumber
\mathcal{H} 
            &\equiv \bar{\gamma}^{ij}\bar{D}_i\bar{D}_j e^\phi - \frac{e^{\phi}}{8}\bar{R}
            + \frac{e^{5\phi}}{8}\tilde{A}_{ij}\tilde{A}^{ij} - \frac{e^{5\phi}}{12}K^2 + 2\pi e^{5\phi}\rho \\
            &= 0 
            \label{hamconst}
\end{align}
and
\begin{equation}
\mathcal{M}^i  \equiv \bar{D}_j (e^{6\phi}\tilde{A}^{ij}) - \frac{2}{3}e^{6\phi}\bar{D}^i K - 8\pi e^{10\phi}S^i = 0 \,.
\label{momconst}
\end{equation}
In order to quantify the amount of constraint violation relative to the energy scales involved in the problem, we compare the calculated constraint violation to the root sum of squares of the individual terms in each respective constraint equation - the ``scale" of the constraint. 
We label these two denominators $\left[\mathcal{H}\right]$ and $\left[\mathcal{M}\right]$, with $\mathcal{M} = |\mathcal{M}^i|$.

Alongside the metric, we evolve a pressureless fluid describing the matter content of the universe.
Our code was written to utilize a flux-conservative form of the relativistic hydrodynamic equations \cite{Font:2000pp}; however, here we are primarily interested in results for a $w=0$ cosmological fluid with rest-mass density $\rho_0$ with no initial velocity component.  The equation of motion for the matter fluid in the absence of any initial velocity is then a simple conservation law, allowing us to optimize the fluid code to model a static source,
\begin{equation}
\partial_t \tilde{D} = \partial_{t}(\gamma^{1/2}\rho_0) = 0.
\end{equation}
The contributions of the matter components to the source terms are then $\rho = \rho_0 + \rho_\Lambda$, $S_i = 0$, $S_{ij} = -\gamma_{ij}\rho_\Lambda$, and $S = -3\rho_\Lambda$.

We fix physical scales by setting our grid side length to $L = n H_I^{-1} = N \Delta x$, where $n$ is an arbitrary fraction and $H_I$ is the Hubble scale of an FLRW solution on the initial slice.
Combining this with the number of points per grid side, $N$, allows us to determine the grid spacing $\Delta x$.
In programming units, we work in units of the Hubble scale of an analogous FLRW solution at the time the initial slice is set. 
This also fixes the density of the FLRW solution to be $\rho_{FLRW} = \left(3/8\pi\right)\left(n/N\right)^2$.

As in almost all simulations, numerical inaccuracies accumulate over time leading to substantial growth of constraint violation at late times.
In order to alleviate this problem, we have included a diffusive term in the evolution equations as suggested by \cite{Duez:2004uh}, with varying degrees of strength.  These modifications take the form
\begin{eqnarray}
\label{dampingterms}
\partial_t \phi & = & ... + 0.1 c_H \Delta t \mathcal{H} \\
\partial_t \bar{\gamma}_{ij} & = & ... + 0.5 c_H \bar{\gamma}_{ij} \Delta t \mathcal{H} \\
\partial_t \tilde{A}_{ij} & = & ... - 1.0 c_H \tilde{A}_{ij} \Delta t \mathcal{H}.
\end{eqnarray}

These additional terms do suppress constraint violation at some level, however late-time growth remains large, ostensibly due to the matter coupling. 
Although we do not explore this idea thoroughly, we do present some preliminary analysis in Sec.~\ref{standardcodetests}.  Because of this instability, and because we are interested in percent-level effects, we generally restrict ourselves to timescales on which $\mathcal{H}/[\mathcal{H}] \lesssim 10^{-4}$, and $\mathcal{M}/[\mathcal{M}] \lesssim 10^{-2}$.

To describe the statistical behavior of fields as they evolve, we commonly compute volume-weighted (conformal) averages
\begin{equation}
\bar{f} \equiv \frac{\int\mathrm{d}x \sqrt{\gamma} f}{\int\mathrm{d}x \sqrt{\gamma}} \simeq \frac{\sum_{x_i} e^{6\phi_i} f_i}{\sum_{x_i} e^{6\phi_i}}
\end{equation}
and volume-weighted standard deviations
\begin{equation}
\sigma_f \equiv \sqrt { \frac{N}{N-1} \frac{ \sum_{x_i} e^{6\phi(x_i)} (f(x_i) - \bar{f})^2 }{ \sum_{x_i} e^{6\phi(x_i)} } }
\end{equation}
of various fields ($f$) on spatial slices.

\section{Cosmological Initial Conditions}
\label{ics}

The initial surface from which we evolve should satisfy the Hamiltonian and momentum constraint equations. 
In the case of an FLRW solution, the Hamiltonian equation is one of the Friedman equations, and all terms in the momentum constraint equation are zero.
We are interested in solutions similar to this, but we also wish to examine a universe with non-uniform matter density.
We do not attempt to specify a matter distribution that is perfectly analogous to our universe, but instead restrict ourselves to initial conditions that are easier to explore in the context of the BSSN formalism.

The Hamiltonian and momentum constraint equations by themselves do not uniquely specify a metric; more degrees of freedom exist than can be specified by the constraint equations.
Additionally, even when using choices that simplify the equations (such as in the conformal transverse-traceless decomposition) the equations can still be difficult to solve.
To this end, we relax the requirement that we need to specify an exact matter source for initial conditions, instead specifying the conformal factor $\phi$ itself.
We specify a slice with constant extrinsic curvature whose value is approximately determined by the average matter density, and fluctuations in density and $\phi$ set by a power spectrum. 
At large scales (small $k$) we allow the power spectrum to scale as $P_k \sim k^1$, and at small scales (large $k$) as $P_k \sim k^{-3}$ \cite{Tegmark:2003ud}.  
Given a peak scale $k_*$ and peak amplitude at this scale $P_*$, we use the power spectrum
\begin{equation}
\label{matterpower}
P_k = \frac{4 P_*}{3} \frac{k/k_*}{1+\frac{1}{3}(k/k_*)^4}\,.
\end{equation}

\begin{figure}[htb]
  \centering
    \includegraphics[width=0.45\textwidth]{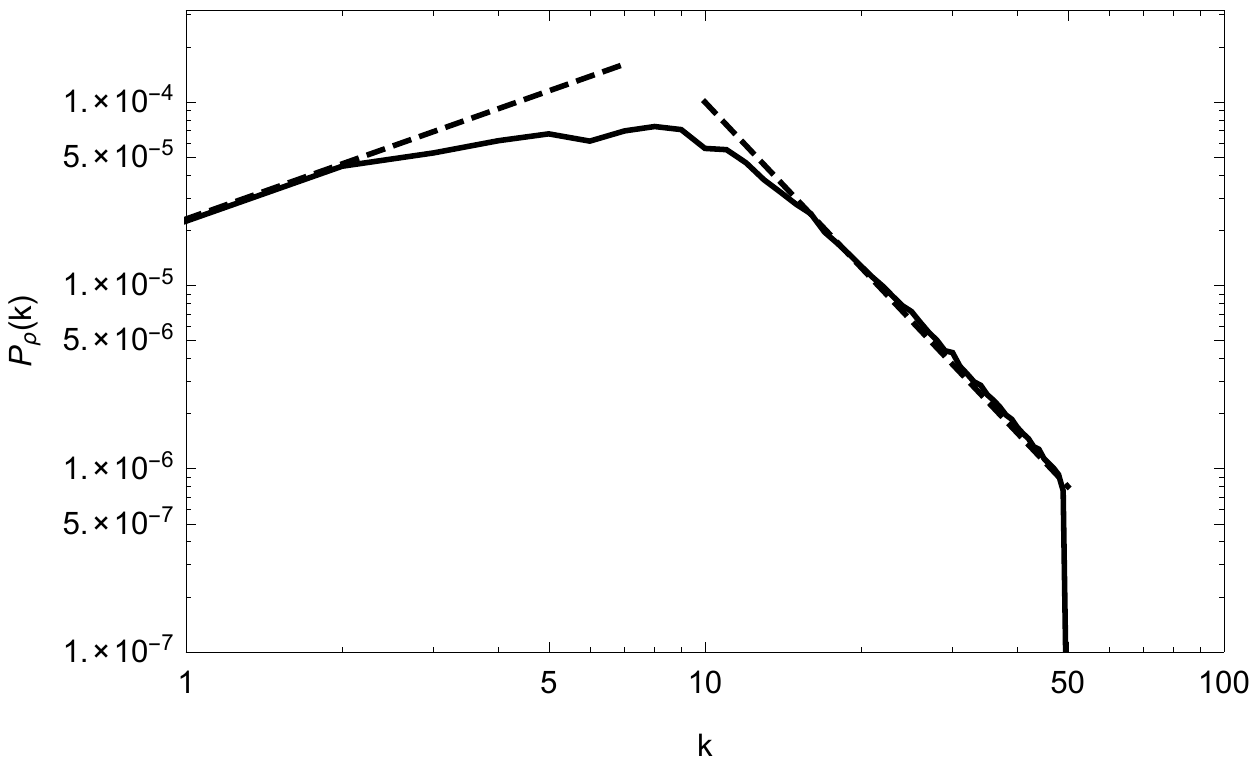}
  \caption{\label{matter_spec}
    Initial matter power spectrum (solid) for an $N^3 = 128^3$ grid, including lines (dashed) indicating $k$ and $k^{-3}$ scalings. The power spectrum is cut at $k = 50 \Delta k$.}
\end{figure}

In order to set these initial conditions, we decompose $\rho$ into two pieces, $\rho_K$ sourcing the trace of the extrinsic curvature $K$ and $\rho_\psi$ sourcing the conformal factor $\psi \equiv e^\phi$, so the total density is $\rho = \rho_K + \rho_\psi$.
We also impose that the matter is at rest, so $S^i = 0$.
We use a conformally flat metric and set the trace-free part of the extrinsic curvature to zero, leaving us to solve two simpler equations:
\begin{eqnarray}
\label{matter_ICs}
\nabla^{2}\psi & = & -2\pi\psi^{5}\rho_{\psi}\\
K & = & -\sqrt{24\pi\rho_{K}}\,.
\end{eqnarray}
The nonlinear $\psi$-equation is difficult to solve for a fixed matter source $\rho_\psi$, with attempted relaxation and iterative solution methods tending to find the $\psi=0$ solution.
We instead opt to specify initial conditions by setting $\psi$ directly, with a power spectrum $P_k^\psi = k^{-4} P_k$. We set the monopole term to unity, and subsequently set $\rho_{\psi} = \nabla^{2}\psi/(-2\pi\psi^{5})$. 
A power spectrum realization can be seen in Fig.~\ref{matter_spec}.

We note that because we cannot directly specify that $\rho_\psi$ has zero mean, the extrinsic curvature $K$ will not necessarily be set by the average of the total density $\bar{\rho}$.

We generate a Gaussian random field using the matter power spectrum in Eq.~\ref{matterpower}, specifying a unitary monopole.
The field $\rho_\psi$ is then calculated using finite differencing from Eq.~\ref{matter_ICs}.
When the Hamiltonian constraint is subsequently calculated, this means the predominant error should only be due to finite precision.
The matter field is constructed as $\rho = \psi^{5}\rho_{\psi} + \rho_K + \rho_\Lambda$.
Lastly, the trace of the extrinsic curvature is set to $K = -\sqrt{24\pi(\rho_K + \rho_\Lambda)}$.

\section{Numerical Method}
\label{compute}

\subsection{Code Structure}
Here we present details of the code used to evolve the BSSN equations.  
The engine of the code is written in {\sc c++} and employs several {\sc c++11} standard features, along with {\sc OpenMP} \cite{dagum1998openmp} for parallelization, FFTW \cite{FFTW05} for calculating power spectra and creating initial conditions, and the HDF5 file format \cite{hdf5} for storing samples of data.  
The simulations presented here are performed on resources available at Kenyon College \cite{tomweb} and the High Performance Computing System at Case Western Reserve, on single nodes that range between 8-64 cores and 64-512 GB of RAM.

The code is divided into different classes each of which evolves a different component. There are two main classes: one that solves the gravitational equations and one that evolves the hydrodynamical equations that define the matter content.  There are three additional classes that implement the static $w=0$ perfect fluid examined here, a class that defines a cosmological constant, and another that can integrate the pure Friedmann equations for a perfect fluid.

The most important of these is the class that evolves the gravitational equations; this ``BSSN class" provides functionality to evolve the BSSN fields in the presence of an arbitrary matter source using a standard fourth-order Runge-Kutta integrator, with four memory registers per gravitational field and an additional register for storing source terms.  We informally note that this code runs at a speed comparable to ETK/CACTUS \cite{EinsteinToolkit:web,Loffler:2011ay}, but seems to consume roughly an order of magnitude less memory.  The BSSN class also includes routines for performing computations such as calculating average momentum and Hamiltonian constraint violation. The results we present are almost exclusively in synchronous gauge, and as such the code has been optimized to take advantage of its simplicity. However, the BSSN class is still capable of evolving the metric in an arbitrary gauge.

The other major class is the ``Hydrodynamics class," which allows us to evolve a perfect fluid with a cosmological equation of state. This class has been simplified for a static $w=0$ fluid, with further work needed to implement an arbitrary equation of state. As either an N-body code or special lattice methods are required to evolve a pressureless fluid with a non-zero velocity component, we constrain ourselves to work with initial conditions where the fluid is static.  In the chosen gauge, the equations of motion simplify such that the conserved fluid density does not evolve, $\partial_t \tilde{D} = 0$.

Finite-difference stencils are implemented up to eighth-order, as numerical stability and accuracy is highly dependent on structure at smaller scales, as demonstrated below. Unless otherwise specified, the stencils used are generally $\mathcal{O}(\Delta x^6)$.

The simulations have periodic boundary conditions, imposed at the level of an array indexing function. However, alternate boundary conditions can be implemented.  With our code we expect to be able to run for an arbitrarily large grid---although it is currently limited to single-node operation.  We have been able to run at resolutions up to $N^3 = 768^3$ on a single 512 GB RAM node.

\subsection{Standard Code Tests: Analytic Cases}
\label{standardcodetests}

We present several code tests where we evaluate several known, analytic solutions, intended to examine the behavior of fields that will dominate the dynamics in cosmological scenarios. The intention is to  examine the validity of results presented, and to explore in what regimes the code will produce accurate and interesting results.

We begin with two standard ``Apples with Apples" tests \cite{Babiuc:2007vr}, the robust stability test and linearized wave test.

\subsubsection{Robust Stability Test}

\begin{figure}
  \centering
    \includegraphics[width=0.45\textwidth]{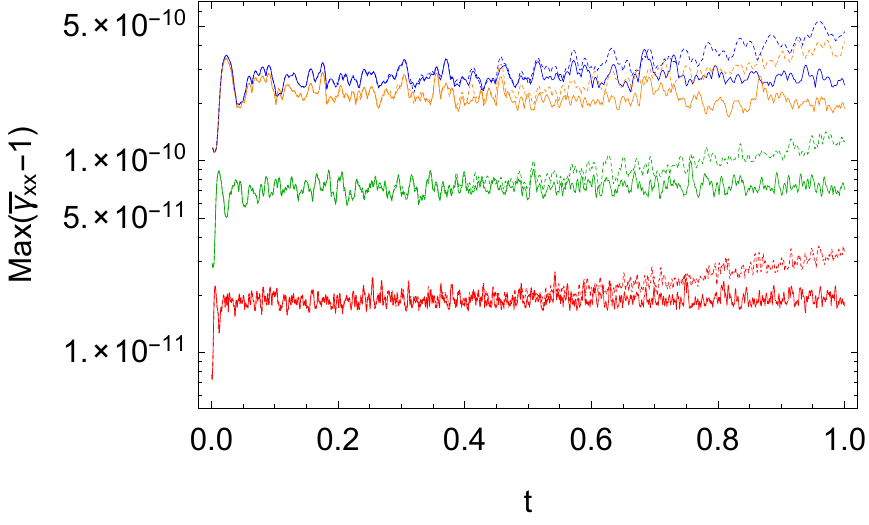}
    \includegraphics[width=0.45\textwidth]{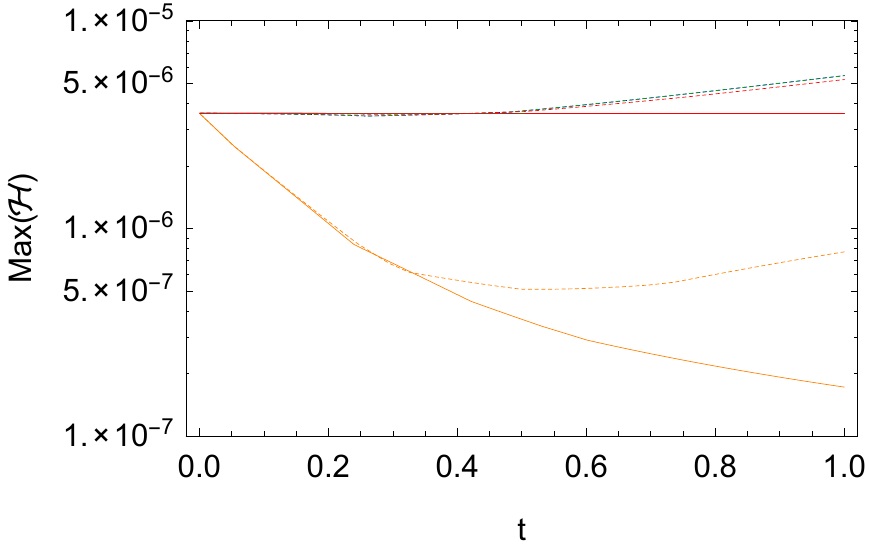}
  \caption{\label{robust_plot_1}
    The top panel shows deviation from Minkowski space versus time for varying $\rho_{\rm AwA}$ for one box crossing time. Blue and orange curves (topmost) correspond to $\rho_{\rm AwA} = 1$, green (middle) to $\rho_{\rm AwA} = 2$, and red (bottom) to $\rho_{\rm AwA} = 4$. Dashed lines include random fluctuations around $\tilde{D} = 0$. The orange curves include damping with $c_H = 1$. At later times, growth of deviations from Minkowski can be seen when noise in the matter sector is included.  The bottom panel shows the maximum constrain violation vs time, with colors as in the top figure. The matter coupling can be seen to introduce a growing mode; the fixed error present in the static $\tilde{D}$ field will source the metric continuously. The constraint damping term helps diminish constraint violation short-term.}
\end{figure}

The robust stability test examines the growth of small amounts of numerical error by looking at the evolution of noise around a Minkowski background in a simulation with no matter content.  Fig.~\ref{robust_plot_1} displays this deviation from Minkowski space with different resolutions.  In this test, we use an asymmetric box in which there are only 6 points in the $y$- and $z$-directions and a variable number of points in the $x$-direction.  In \cite{Babiuc:2007vr}, the number of points in the $x$-direction is parameterized by $\rho_{\rm AwA}$ (the subscript here is added to avoid confusion with the density parameter) such that the number of points in the $x$-direction is $N_x = 50\rho_{\rm AwA}$.

\subsubsection{Linearized Wave Test}

\begin{figure}[htb]
  \centering
    \includegraphics[width=0.45\textwidth]{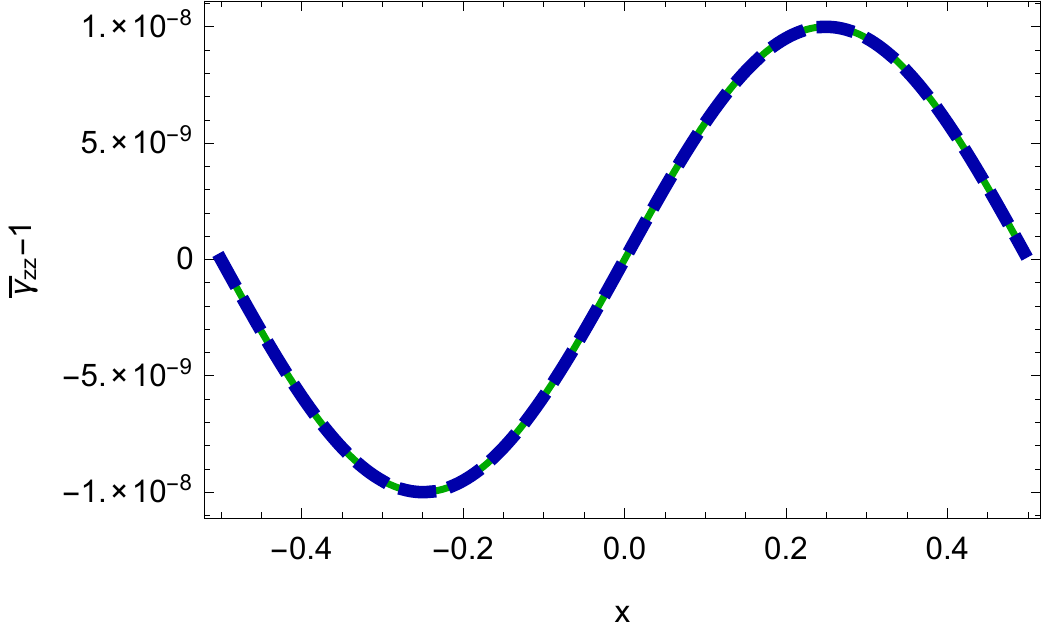}
    \includegraphics[width=0.45\textwidth]{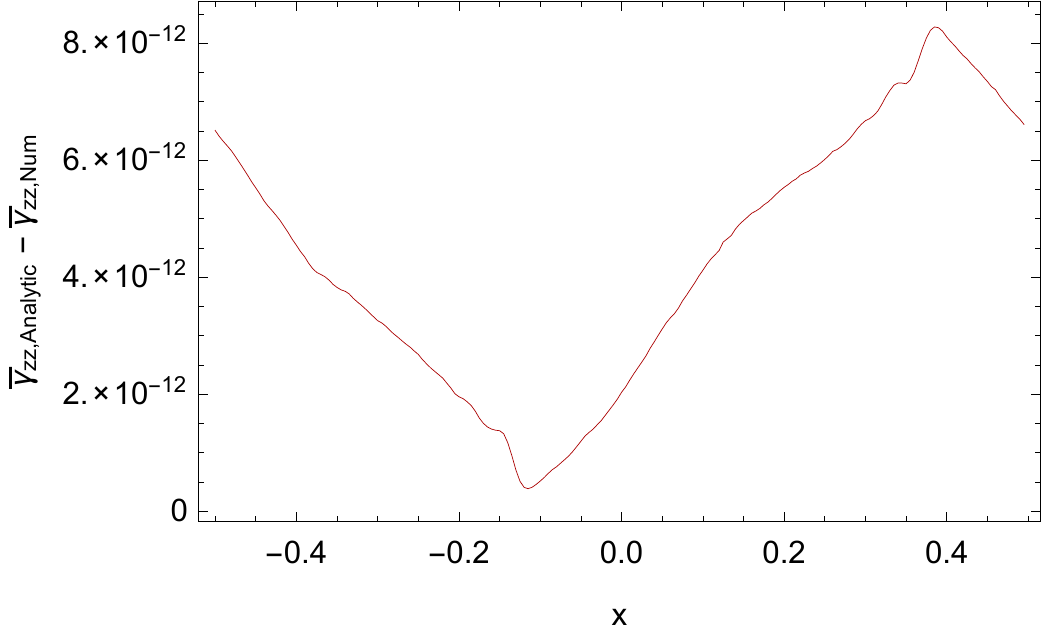}
  \caption{\label{linear_wave_test_1}
    The top panel shows the analytic (blue, solid) and numerical (red, dashed) solutions for the linear wave test after 1000 box crossing times for $\rho_{\rm AwA} = 4$.  The bottom panel shows the difference between the two curves in the top panel, scaled so differences are noticeable.}
\end{figure}

The linearized wave test examines the behavior of linear metric fluctuations after a large number of box light-crossing times. Since we have a plane wave, we work with the asymmetric box of the previous section; where $N_y=N_z=6$ and $N_x = 50\rho_{\rm AwA}$.  Spacetime is initialized with plane wave traveling in the $x$-direction---the direction with the largest resolution---with an amplitude small enough that nonlinear effects are at the level of roundoff error.  Fig.~\ref{linear_wave_test_1} shows the difference between the numerical result of this wave compared to the expectation from linearized gravity.

\subsubsection{Black Hole Profile}

\begin{figure}[htb]
  \centering
    \includegraphics[width=0.45\textwidth]{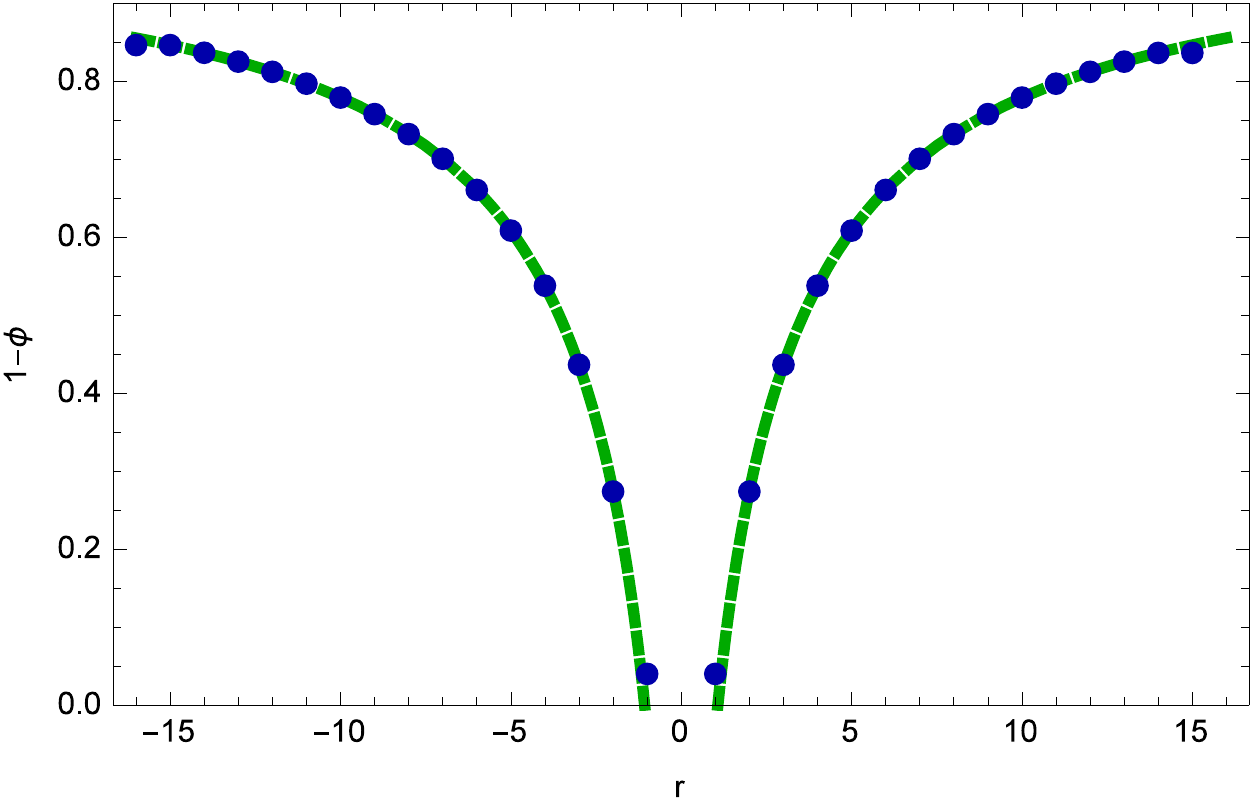}
  \caption{\label{bh_plot}
    A cross section of the conformal field $\phi$ after a period of relaxation, analytically (dashed green line) and numerically (blue dots) for a black hole.
  }
\end{figure}

To verify the validity of the code for fluctuations in the conformal field $\phi$ and gauge fields, we examine an analytic black hole solution with a ``trumpet" geometry \cite{Baumgarte:2007ht} in the `$1+\log$' slicing. The solution is stable at remarkably low resolution, $N^3 = 32^3$, and with second-order finite-derivative stencils.
After initializing our grid with the solution, we see it relax slightly, and subsequently remain stable.  
This solution is run in vacuum, using a time-independent maximal slicing gauge condition for the lapse, and hyperbolic Gamma driver shift condition \cite{Campanelli:2005dd}.
At the outer boundary we na\"ively apply fixed boundary conditions.
In the interior of the black hole the lapse approaches zero, so the value of the conformal field $\phi$ is set such that the derivatives of $\phi$ at nearby points will be calculated to a good approximation.
We plot the conformal field, $\phi$, in the simulation and compare it to the analytic black hole solution in Fig.~\ref{bh_plot}.

\subsubsection{FRW universe}

\begin{figure}[htb]
  \centering
    \includegraphics[width=0.45\textwidth]{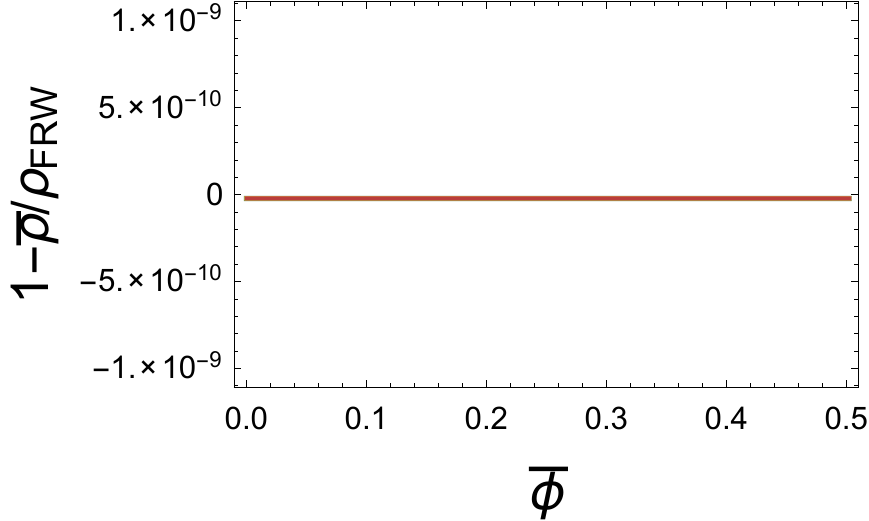}
    \includegraphics[width=0.45\textwidth]{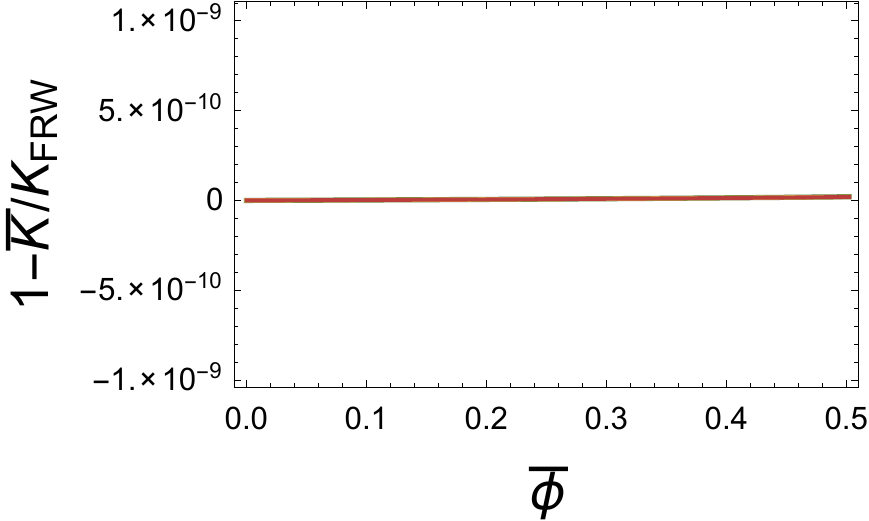}
  \caption{\label{frw_plots}
    Deviation of FRW quantities from analytic solutions for varying timesteps, and resulting fractional constraint violation. The colors red, yellow, and green correspond to timesteps $\Delta t = \Delta x/10$, $\Delta t = \Delta x/20$, and $\Delta t = \Delta x/40$ respectively. Here $\Delta x = L/N = H_I^{-1}/2N$.
    }
\end{figure}

We also examine the case of a homogeneous flat FLRW universe. 
For this we simply choose geodesic slicing, or synchronous gauge, with periodic boundary conditions on a very small ($N^3 = 8^3$) grid.  
Since there are no spatial inhomogeneities in the simulation, spatial resolution should not play a role in its validity.
Indeed, Fig.~\ref{frw_plots} shows that the software can reproduce FLRW cosmology to high accuracy.

\subsection{Cosmological Code Tests: Variation of Parameters}

To test our code in a cosmological regime, we define a fiducial model and then vary physical and numerical parameters to test the stability and reliability of our simulations.
The fiducial model has $N^3 = 128^3$ points on each side; each side of the simulation is half the corresponding FLRW Hubble scale, $L = H_I^{-1}/2$. 
We employ $\mathcal{O}(\Delta x^6)$ finite-derivative stencils, and $\Delta t = \Delta x/10$.
The initial conditions have a power spectrum cut at $k_{\rm cutoff} = 10/128\,\Delta k$, peak frequency of $k \sim 7/128\,\Delta k$, and peak amplitude such that $\sigma_\rho / \rho \sim 0.04$.
We note that the resolution we have chosen corresponds to $\Delta x\sim 16$ MPc, somewhat close to the 11.5 MPc $\sigma_8$ scale. 
However, even without the cut, the amplitude corresponds to $\sigma_\rho / \rho \sim 0.1$. Thus a combination of the spectrum cutoff and smaller spectrum amplitude result in relatively small amplitude fluctuations compared to the measured value $\sigma_8 \sim 0.8$ today \cite{Ade:2015xua}; thus we are studying a smoother universe.

\begin{figure}[htb]
\centering
\includegraphics[width=0.45\textwidth]{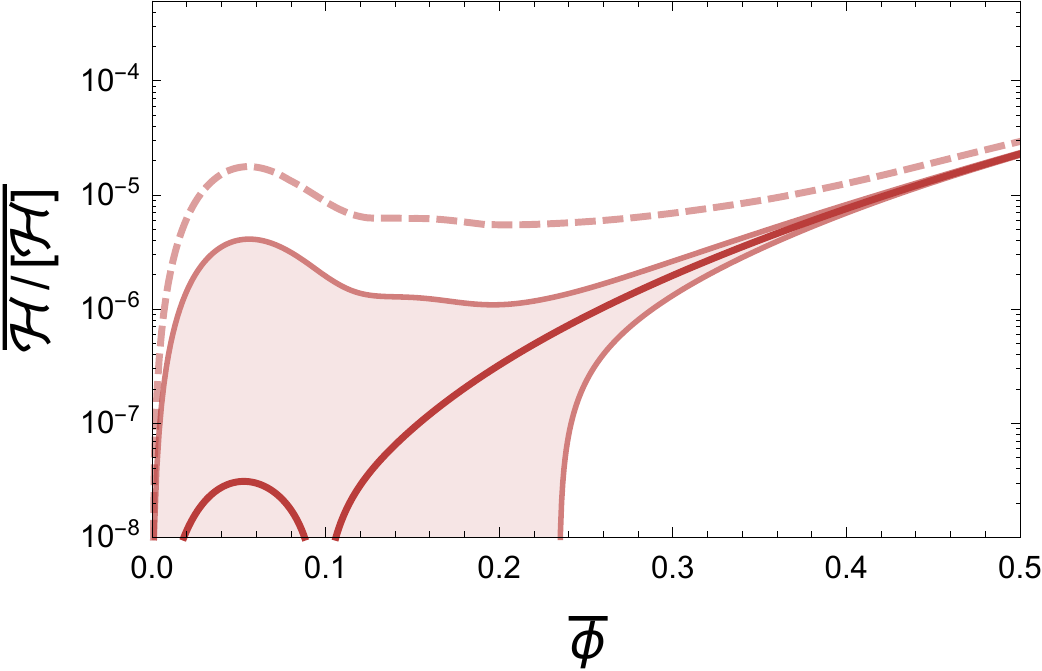}
\includegraphics[width=0.45\textwidth]{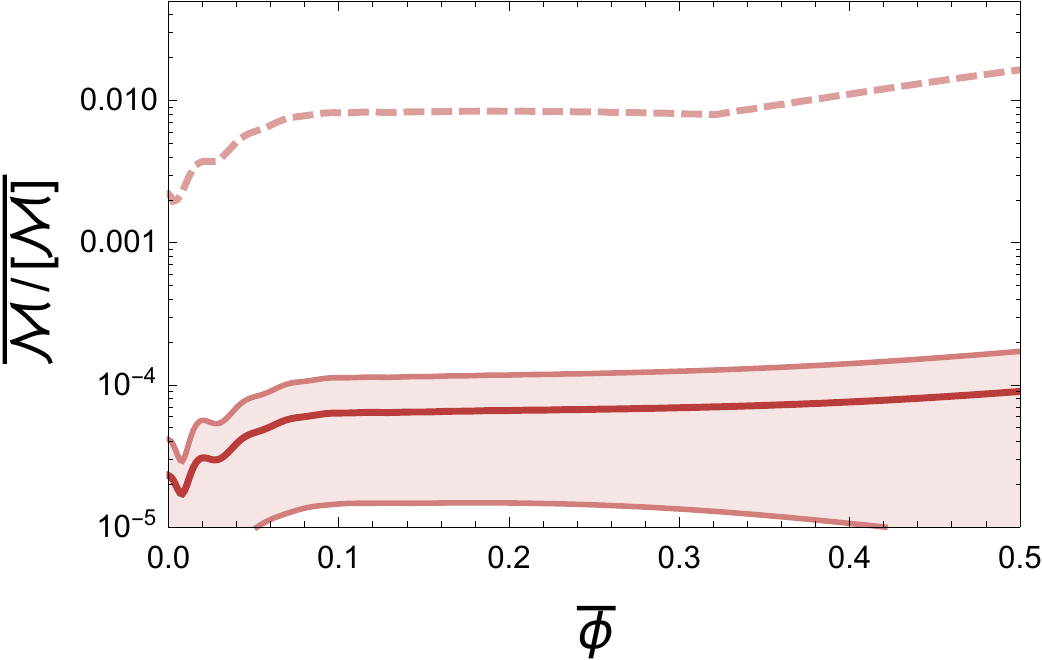}
\caption{\label{fiducial_constraints} We show the satisfaction of the constraints, Hamiltonian (left panel) and momentum (right panel), for our fiducial model.  The bold lines show the average value of the constraint, the shaded regions envelop 68\% of the points of the grid and the dashed line shows the worst (most violation) point.}
\end{figure}

\begin{figure}[htb]
\centering
\includegraphics[width=0.45\textwidth]{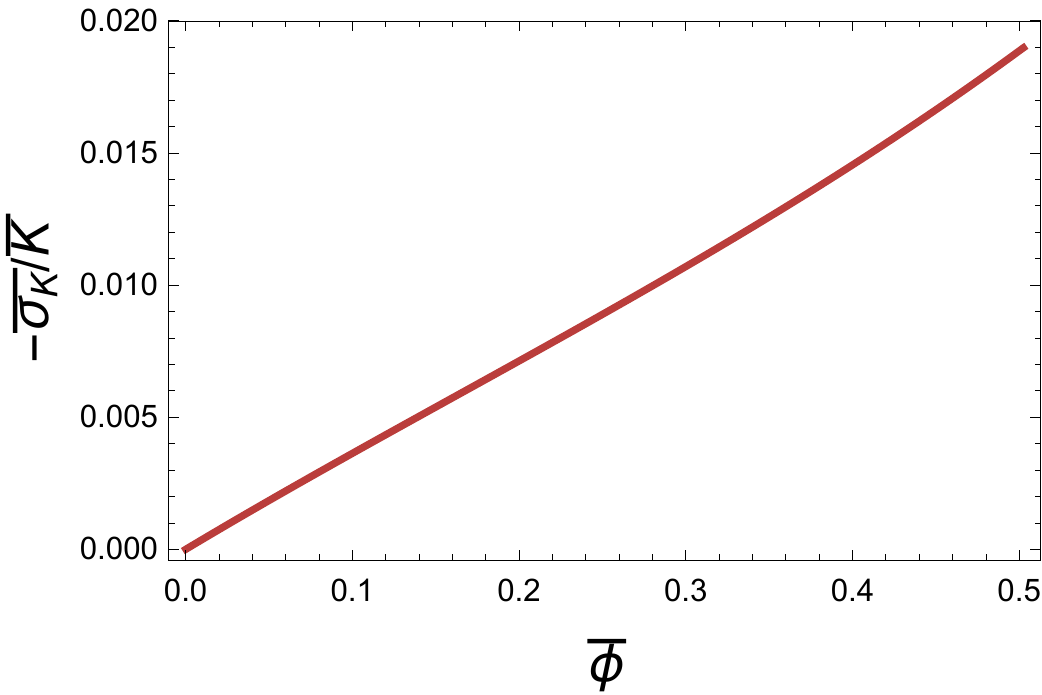}
\caption{\label{fiducial_results} The level of inhomogeneity in the extrinsic curvature, $\sigma_K/\bar{K}$ as a function of the average conformal factor, $\bar{\phi}$, for our fiducial model.}
\end{figure}

So that we have a method of comparison, we present some baseline information about our fiducial model. Fig.~\ref{fiducial_constraints} shows how well our fiducial model satisfies the physical constraints, Eqs.~\ref{hamconst} and \ref{momconst}, and Fig.~\ref{fiducial_results} shows how inhomogeneities of the extrinsic curvature, $\sigma_K/\bar{K}$, are generated over the course of the simulation.

\subsubsection{Constraint Violation}

\begin{figure}[htb]
  \centering
    \includegraphics[width=0.45\textwidth]{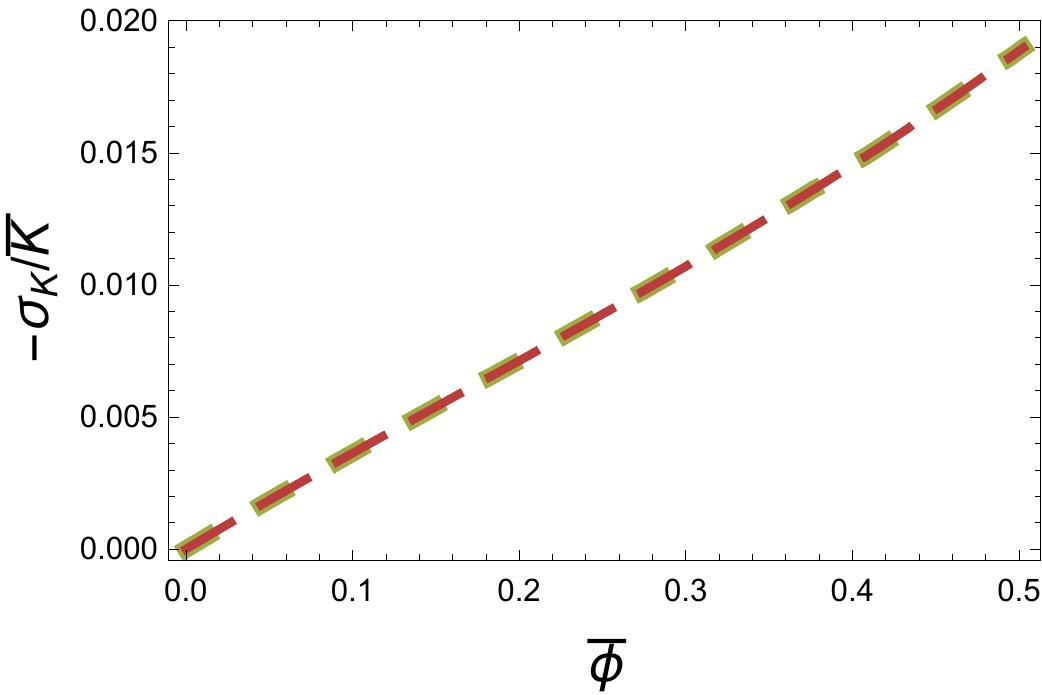}
    \includegraphics[width=0.45\textwidth]{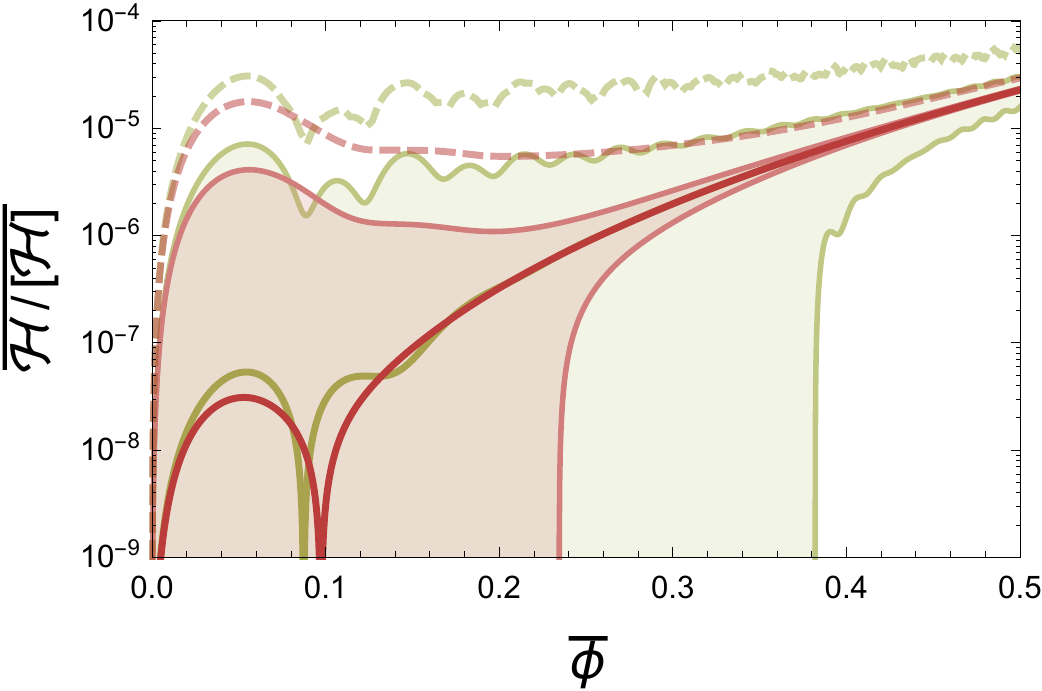}
  \caption{\label{constraint_variaion_damping}
We vary the value of $c_H$ in Eq.~\ref{dampingterms}.  The top panel shows the effect of varying this parameter on the satisfaction of the Hamiltonian constraint, Eq.~\ref{hamconst}, over time.  The bottom panel shows the level of Hamiltonian constraint violation as we vary the same parameter.  In both panels we take $c_H$ to be $0$ (green) and $40$ (red).   The bold lines show the average value of the constraint, the shaded regions envelop 68\% of the points of the grid and the dashed line shows the worst (most violation) point.}
\end{figure}

Here we explore the effect of the constraint damping terms from Eq.~\ref{dampingterms}. Increasing the magnitude of $c_H$ helps reduce the maximum amount of Hamiltonian constraint violation, although it does not have much effect on the overall growing mode.  Fig.~\ref{constraint_variaion_damping} shows the effect of this term on level of Hamiltonian constraint violation and on the generation of inhomogeneities of the extrinsic curvature, $K$.

\subsubsection{Varying the Resolution}
\label{resolutiontest}

\begin{figure}[htb]
  \centering
    \includegraphics[width=0.45\textwidth]{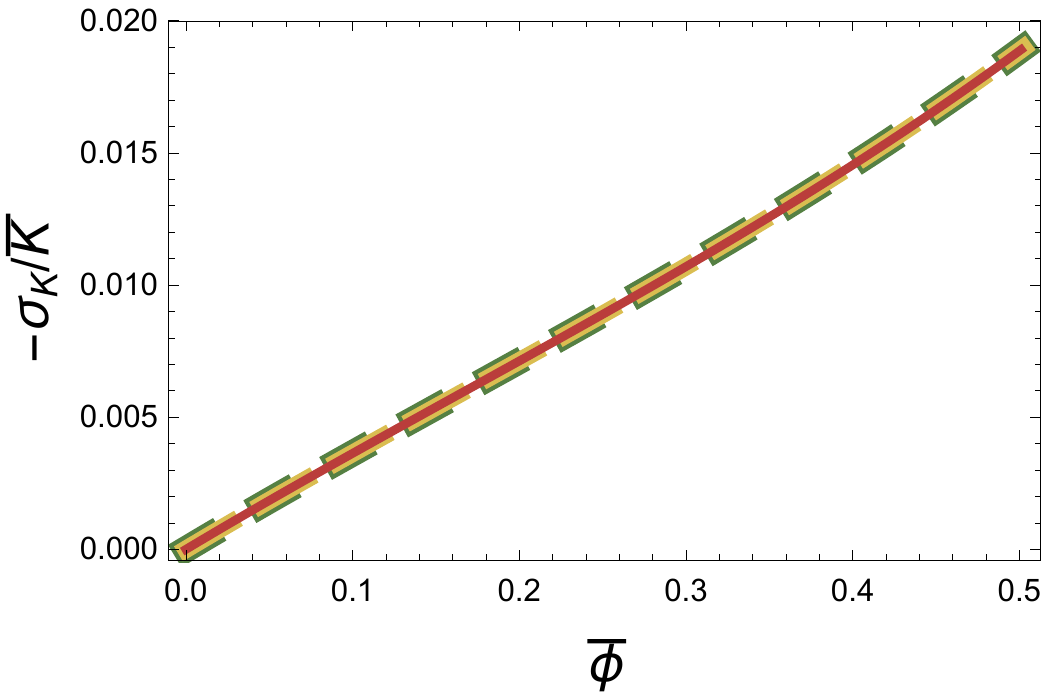}
    \includegraphics[width=0.45\textwidth]{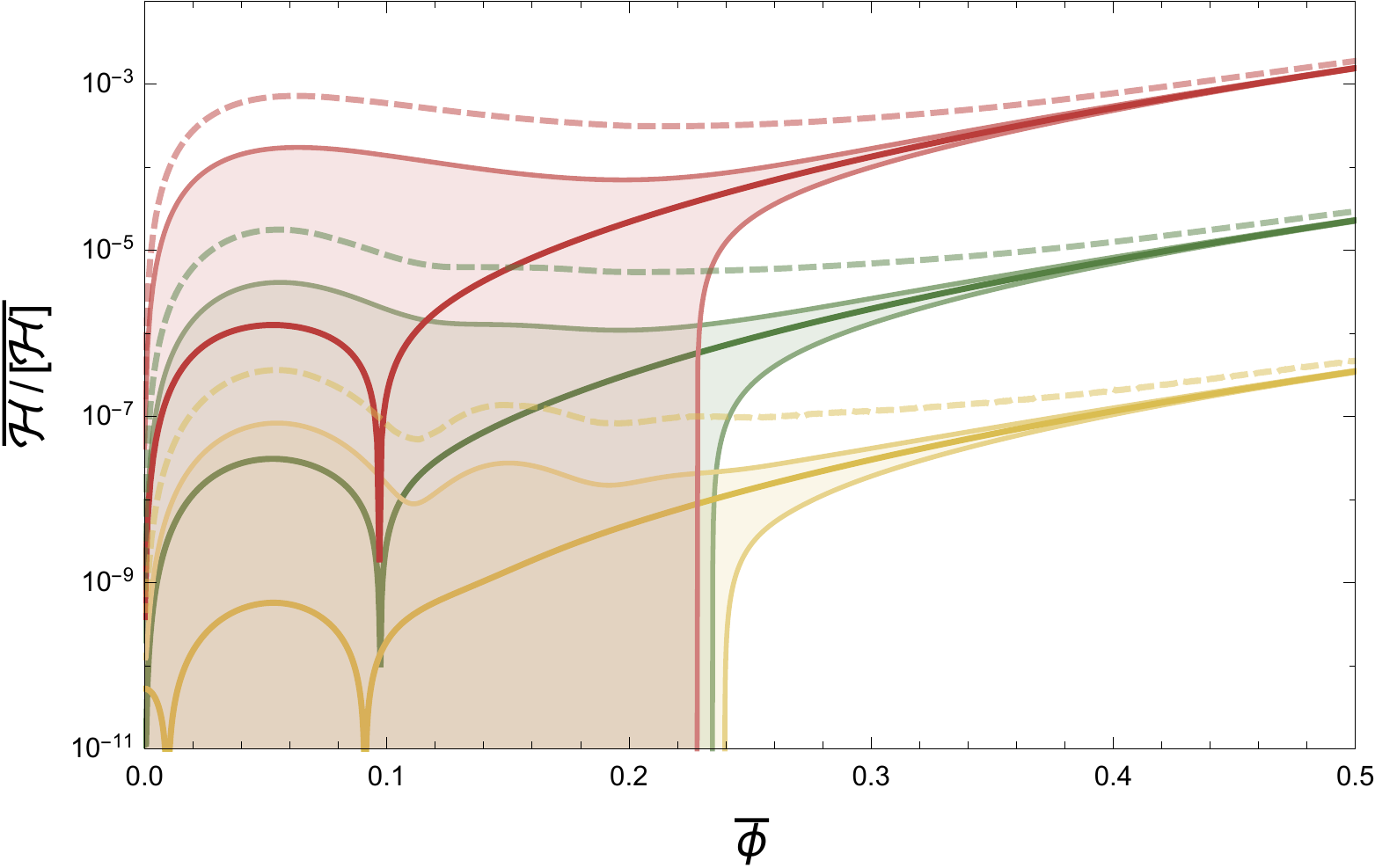}
  \caption{\label{resolution_vary_1}
    We track the generation of inhomogeneities of the extrinsic curvature, $\sigma_K/\bar{K}$, (top panel) and the satisfaction of the Hamiltonian constraint, Eq.~\ref{hamconst}, as we vary the resolution of the simulation.  In both plots the color scheme corresponds to resolutions $64^3$ (red), $128^3$ (fiducial, green), and $256^3$ (yellow).  We see the behavior of $\sigma_K/\bar{K}$ is stable and the amount of constraint violation diminishes as resolution is increased.  The bold lines show the average value of the constraint, the shaded regions envelop 68\% of the points of the grid and the dashed line shows the worst (most violation) point.}
\end{figure}

Here we examine the effects of changing the resolution while keeping the same initial conditions (picking initial conditions with identical amplitudes for $\vec{k}$). We see that the amount of constraint violation diminishes as the resolution is increased, something expected due to the increasing smoothness of the fields, making derivatives more accurate. Fig.~\ref{resolution_vary_1} shows that we see no qualitative difference in the generation of inhomogeneities and marginally better satisfaction of the Hamiltonian constraint, Eq.~\ref{hamconst}, as we increase the resolution of the simulations. 

\begin{figure}[htb]
  \centering
    \includegraphics[width=0.45\textwidth]{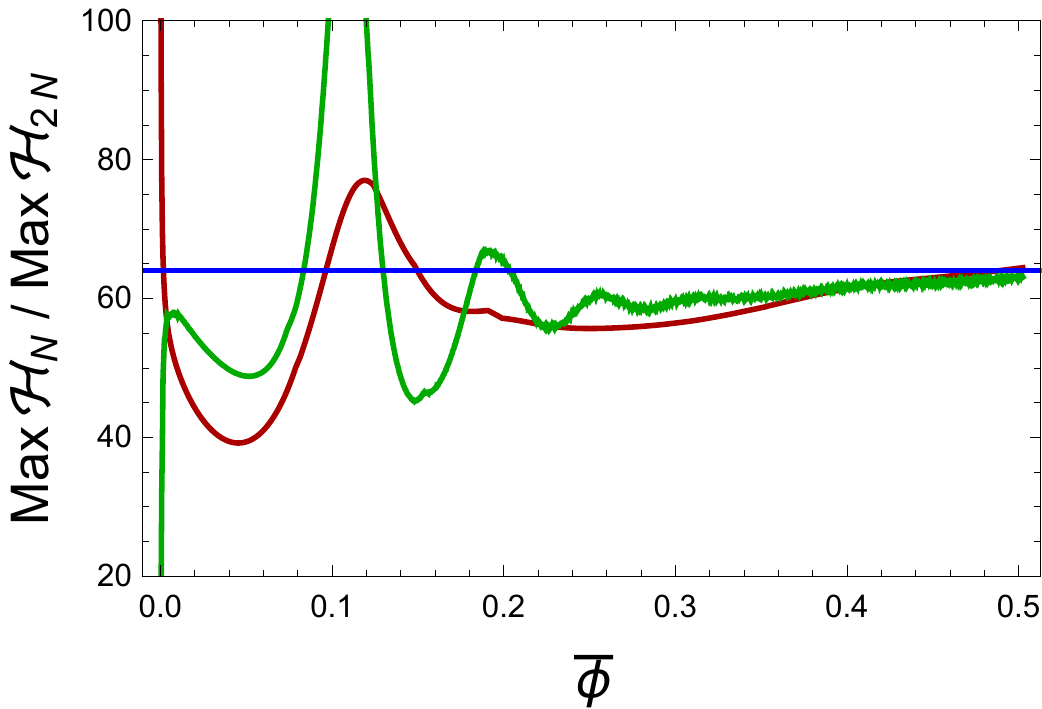}
  \caption{\label{resolution_vary_3}
    Sixth-order convergence can be seen as resolution is varied. The ratio of errors $||H||_\infty$ at two resolutions $\Delta x = L/N$ and $\Delta x' = L/(2N)$ for an $\mathcal{O}(\Delta x^6)$ method should be $2^6 = 64$, plotted as a blue line. Curves for $N^3 = 64^3$ and $N^3 = 128^3$ are shown.
    }
\end{figure}

A more precise test of resolution variation can be seen in Fig.~\ref{resolution_vary_3}.  We should be able to predict how much better the constraints can be satisfied as we increase the resolution; specifically, as we double the number of points on a side, we should see the ratio of errors be $\mathcal{O}(\Delta x^6) / \mathcal{O}(\Delta x^6 / 2^6) \sim 2^6$.

We have implemented a number of finite-differencing stencils, seeing reduced error for higher-order stencils as expected. The use of $\mathcal{O}(dx^2)$ stencils in particular lead to unacceptable growth of constraint violation in our simulations.  We also tested changing the time resolution $\Delta t$ to check for convergence. The results presented here agree with runs for $\Delta t$ an order of magnitude smaller.

\subsubsection{Varying the Spectrum Cutoff}

As a final technical test, we present results as the spectrum cutoff is varied. As the cutoff approaches smaller $\vert\vec{k}\vert$, we see that the amplitude of fluctuations is significantly diminished, as expected, when modes are removed. The relative amount of constraint violation also diminishes as the fields become increasingly smooth. As the cutoff is raised, we see the opposite happen: structure at increasingly small (and unresolvable) scales is created, and constraint violation increases.  The amount of constraint violation after an $e$-fold is a part in ten when $c=40$, and is of order a part in $10^{11}$ for $c = 2$. 
\begin{figure}[htb]
  \centering
    \includegraphics[width=0.45\textwidth]{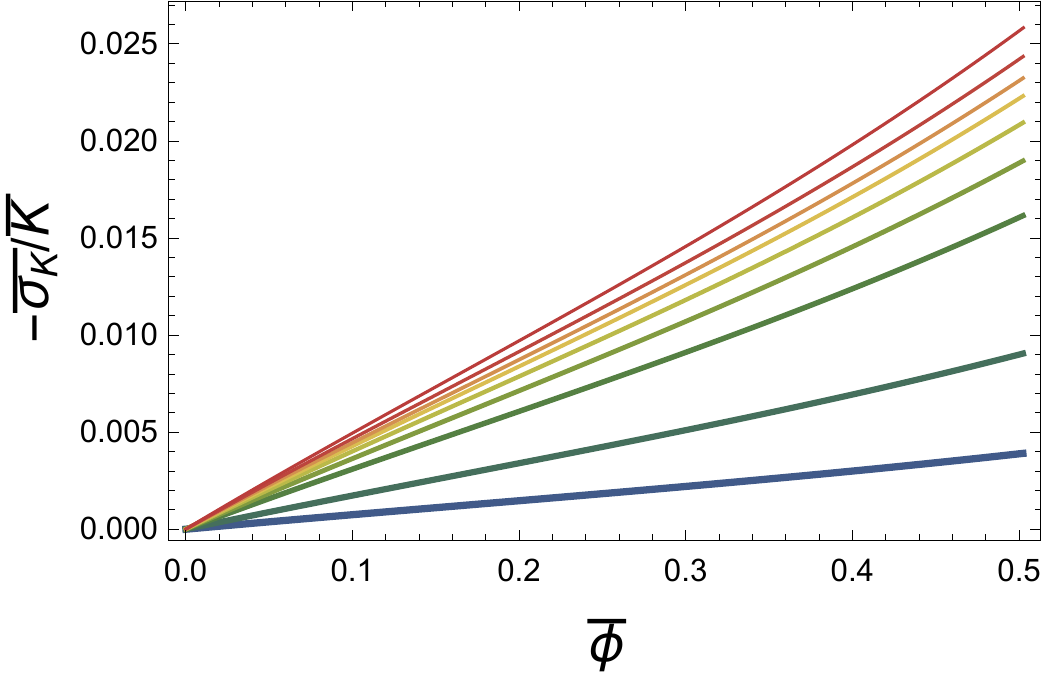}
  \caption{\label{spectrum_vary}
    The generation of inhomogeneities in the extrinsic curvature, $\sigma_K/\bar{K}$ as we vary $k_{\rm cutoff}$.  The color scheme follows $k_{cut} = c \Delta k$, for $c = 2, 4, 8, 10, 12, 14, 16, 20, 40$, from blue to red (bottom to top). Note that as we initialize more modes, we are also increasing $\sigma_\rho/\bar{\rho}$.}
\end{figure}

This is an important test, specifically when compared to the resolution test, Sec.~\ref{resolutiontest}, as it shows that we have control over the finite differencing scheme; for example, doubling the value of $k_{\rm cutoff}$ resolves the same {\sl physical} modes as would halving the physical size of the box (and keeping the resolution constant), or doubling the resolution of the box.  While the {\sl numerical} strategy differs for these three cases by many factors of $\Delta x$, $\Delta t$, $N$, etc. we see excellent agreement in the results from our code.  We have demonstrated that these three situations produce the same physical results and show that we are using our software well within the regime of its validity.

\section{Results}
\label{results}

Now that we have demonstrated the stability and reliability of our simulations, we complement the main results of these simulations as presented in \cite{Giblin:2015vwq}.  We are primarily interested in understanding how inhomogeneous matter generates local differences in the expansion rate.  We do this in two steps.  First, we look to see if we generate fluctuations of the extrinsic curvature, $\sigma_K/\bar{K}$, which, in our gauge, is is the source for the conformal factor, $\phi$.  Once this is established, we will run two tests to tell whether or not this generation is an artifact of our scheme or has a measure that shows local deviation from FLRW.

\begin{figure}[htb]
  \centering
    \includegraphics[width=0.45\textwidth]{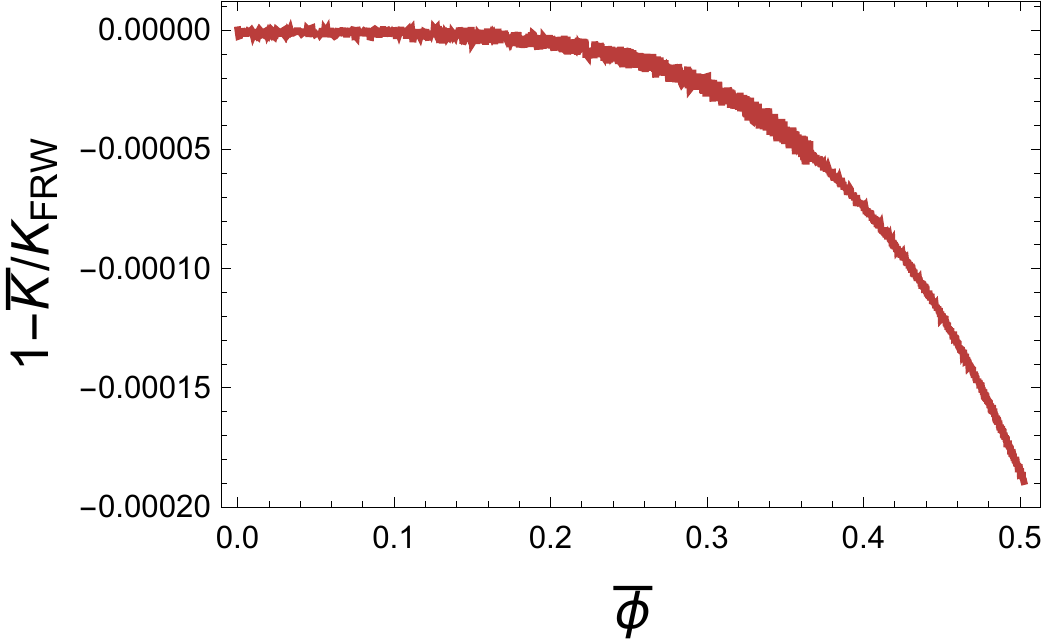}
    \includegraphics[width=0.45\textwidth]{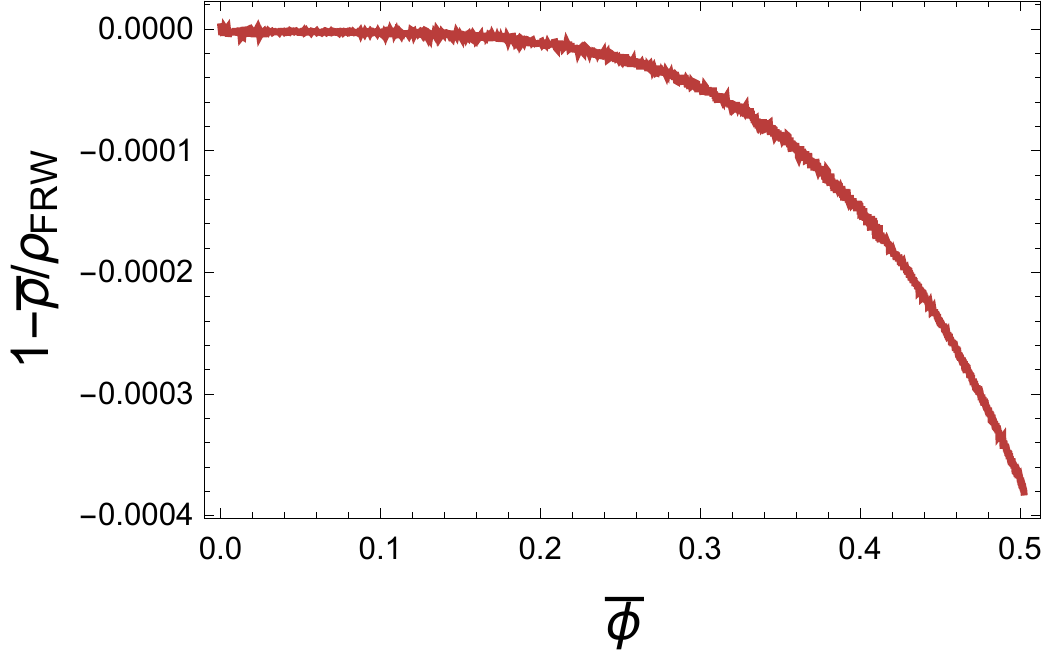}
  \caption{\label{frw_plots_tofid_1}
    Here we show how our fiducial model compares to the numerical FRW models.}
\end{figure}

We compare the fiducial model to the corresponding FLRW simulation with the same initial average initial density, $\bar{\rho}$.  The results are  shown in Figs.~\ref{frw_plots_tofid_1} and show that {\sl average} quantities are very consistent with those in the corresponding homogeneous FLRW universe.
Average parameters correspond very well with FLRW parameters for the $w = 0$ case we study here. 
However, we do begin to see small deviations from the FLRW approximation that are robust to resolution increase. The significance of this result is on the order of the constraint violation. 

On the other hand, we always generate deviations from homogeneity in the extrinsic curvature.  We already presented this result for our fiducial model in Fig.~\ref{fiducial_results}.  As the simulation progresses we see significant local departures from constant $K$.

Of course, we want to explore the relationship between the growth of inhomogeneities of $K$ with increasingly inhomogeneous matter sources.  Varying the amplitude has a corresponding effect on the matter over-density (analogue of $\sigma_8$).
\begin{figure}[htb]
    \centering
      \includegraphics[width=0.45\textwidth]{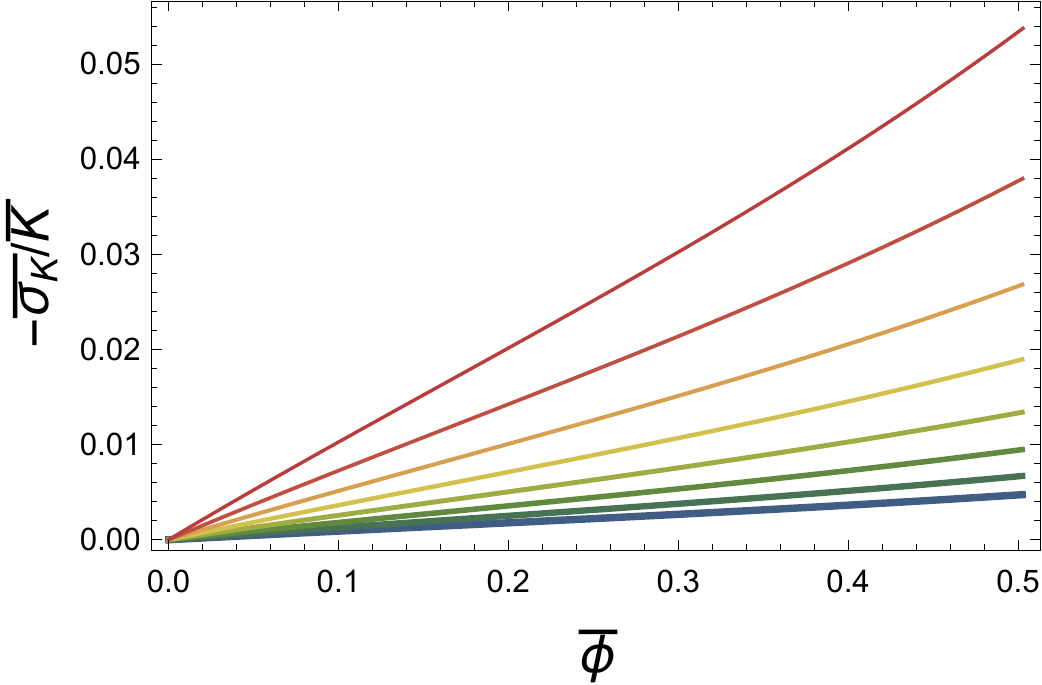}
      \includegraphics[width=0.45\textwidth]{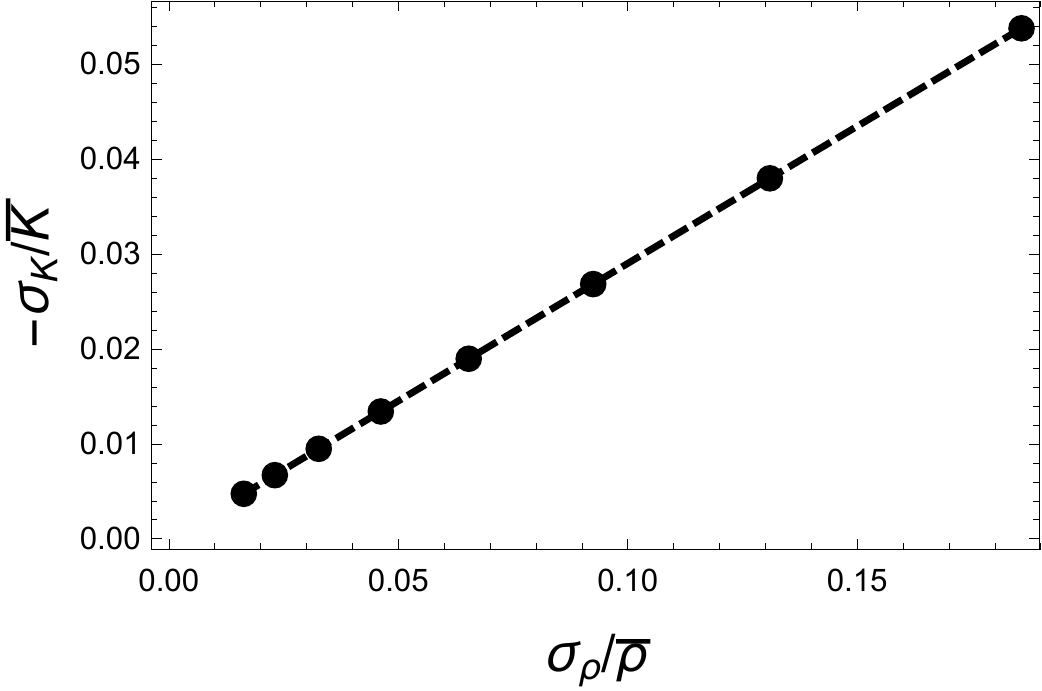}
    \caption{\label{spec_amplitude_1}
The top panel shows variations in the extrinsic curvature, $\sigma_K/\bar{K}$, versus variations in $\sigma_\rho/\bar{\rho}$ over the course of many runs. The initial $\sigma \rho / \bar{\rho}$ for these runs were $\sigma \rho / \bar{\rho} = 0.009$, $0.0133$, $0.019$, $0.027$, $0.038$, $0.053$, $0.076$, and $0.107$ from bottom to top (blue to red).  The bottom panel shows the parametric dependence of $\sigma_K/\bar{K}$ on $\sigma_\rho/\bar{\rho}$ at $\bar{\phi} = 0.5$. }
  \end{figure}

\begin{figure}[htb]
  \centering
    \includegraphics[width=0.45\textwidth]{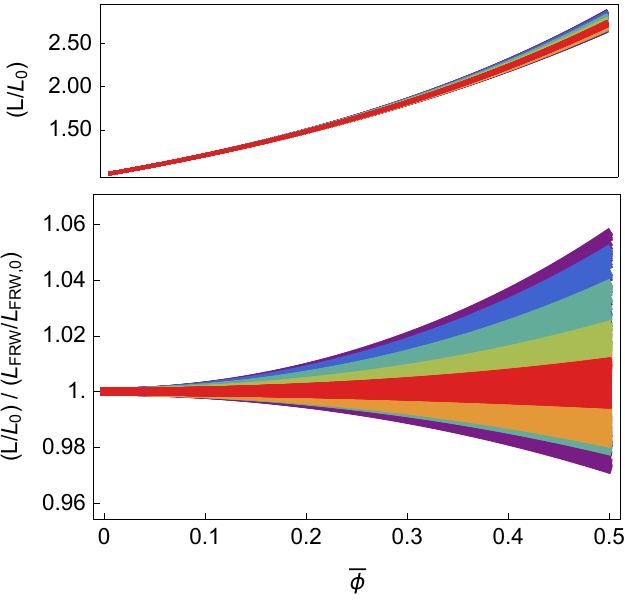}
  \caption{\label{FLRWtest}
The top panel shows the ratio of the proper length to the initial length of six sets of $2^7$ paths in our simulation.  The color coding goes from paths of initial coordinate distance of $4\Delta x$ (purple) to paths of initial coordinate distance $128\Delta x$ (red) in spectral order.  The bottom panel shows the same set of paths, but compares the ratio of proper length to initial proper length with the FLRW estimate.}
\end{figure}

Now that we have established the existence of spatial variations of the extrinsic curvature, $K$, we seek to explore the behavior of additonal measures associated with the spacetime. The first test is to see if there is a departure from FLRW predictions for the proper lengths of paths;  in pure FLRW cosmology, the proper length of {\sl any} path will depend on its initial length and the scale factor, $a_{\rm FLRW}$.  Since we use synchronous gauge, it is a good test to chose arbitrary paths and calculate the ratio of the lengths of these paths as a function of time--if they vary from the FLRW prediction, there is a clear departure from the homogeneous model.  We calculate the length of paths by 
\begin{equation}
S = \int_{\vec{x}_1}^{\vec{x}_2} \sqrt{g_{ij} dx^idx^k}
\end{equation}
between two arbitrary points, $\vec{x}_1$ and $\vec{x}_2$.  Fig.~\ref{FLRWtest} parameterizes the differences between our simulations and the FLRW model for a large set of paths--the paths that we chose are straight coordinate lines.  We note that the departures from FLRW are more significant for shorter paths.  

We also look at the amount of violation of the Killing equation for translational FLRW Killing vectors. The Killing equation reads,
\begin{equation}
\Delta_{\mu \nu} = D_\mu k_\nu + D_\nu k_\mu = 0
\end{equation}
and has solutions $k^\mu$. In a flat FLRW universe, the Killing vectors associated with translations are $(k^\mu_x, k^\mu_y, k^\mu_z) = ( (0,1,0,0), (0,0,1,0), (0,0,0,1) )$. In an inhomogeneous universe, the killing equation should not be perfectly satisfied. We look at the contracted killing equation associated with translations in the $i$-direction,

\begin{equation}
\Delta_i \equiv \Delta_{\mu\,i}^{\mu} = 2 D_\mu k^\mu_i = 2 ( \bar{\Gamma}^j_{ij} + 6 \partial_i\phi ).
\end{equation}

We look at the growth of this quantity relative to the timescale on which the metric evolves, $K$.

\begin{figure}
  \centering
    \includegraphics[width=0.45\textwidth]{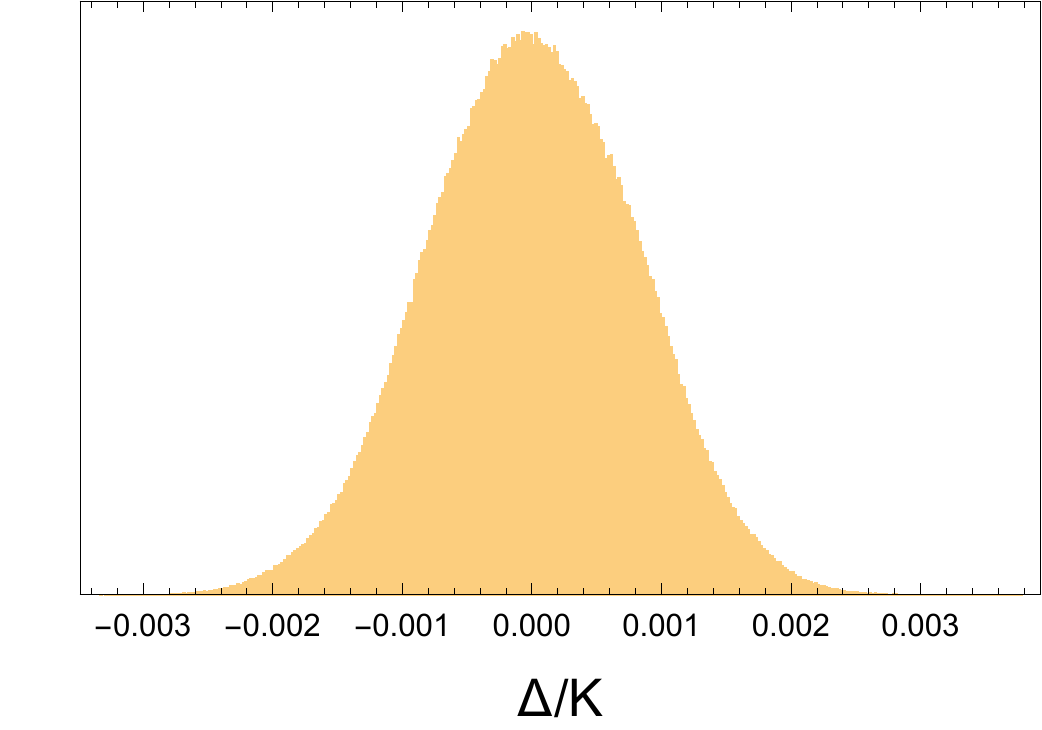}
    \includegraphics[width=0.45\textwidth]{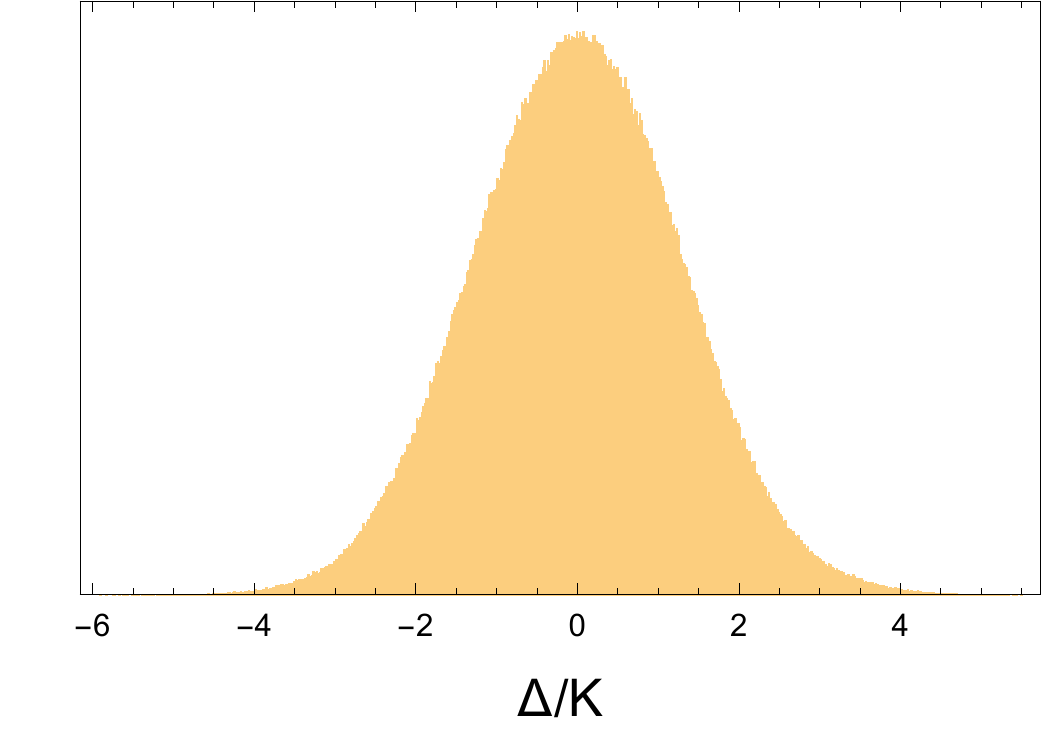}
  \caption{\label{killing_quantity}
    Here we show the distribution (and growth) of $\Delta$, the amount of violation of the Killing equation for the translational FRW killing vectors, relative to the scale $K$ for initial conditions (top) and after half an $e$-fold (bottom).
  }
\end{figure}

\section{Discussion}
\label{discussion}

We have shown that non-linear GR effects are important on sub-Hubble scales in the presence of inhomogeneities in the universe.  Without assuming anything about the dynamical system (e.g. not using symmetry to simplify the dynamical GR equations) we have performed a simulation of the universe that does not rely on a background or averaging procedure.  Using this method we generate FLRW behavior for homogeneous cosmologies and see deviation from this for inhomogeneous ones.  We show that, even beginning with constant extrinsic curvature, $K$, inhomogeneities will generate local fluctuations in this parameter.

Physical observables need to be addressed independently.  For example, the magnitude of the effect on a Hubble diagram would require the integration of photon geodesics during the simulation. We defer this, and other observable tests, for future work.

As the dominant contribution to the BSSN equations in an FLRW spacetime is simply the FLRW solution, it is in principle possible to subtract the FLRW solution from the BSSN equations, and evolve variables representing differences between the FLRW solution and an inhomogeneous solution.  Note that this does not constitute an approximation: the equations are still fully relativistic under this procedure.  We do not explore this idea here, however we may in future work.

\section{Acknowledgments}
We would like to express a special thanks to Thomas Baumgarte for extremely valuable conversations and advice during early stages of this project. JTG would like to thank Eugene Lim and Katy Clough for useful discussions and for the hospitality at King College London at which these conversations took place.  JBM would like to thank Shaden Smith for initial contributions to the code. JBM and GDS are supported by a Department of Energy grant DE-SC0009946 to the particle astrophysics theory group at CWRU. JTG is supported by the National Science Foundation, PHY-1414479. We acknowledge the National Science Foundation, the Research Corporation for Science Advancement and the Kenyon College Department of Physics for providing the hardware used to carry out these simulations. This work made use of the High Performance Computing Resource in the Core Facility for Advanced Research Computing at Case Western Reserve University. 
\FloatBarrier

\bibliography{references}

\begin{thebibliography}{24}
\expandafter\ifx\csname natexlab\endcsname\relax\def\natexlab#1{#1}\fi
\expandafter\ifx\csname bibnamefont\endcsname\relax
  \def\bibnamefont#1{#1}\fi
\expandafter\ifx\csname bibfnamefont\endcsname\relax
  \def\bibfnamefont#1{#1}\fi
\expandafter\ifx\csname citenamefont\endcsname\relax
  \def\citenamefont#1{#1}\fi
\expandafter\ifx\csname url\endcsname\relax
  \def\url#1{\texttt{#1}}\fi
\expandafter\ifx\csname urlprefix\endcsname\relax\def\urlprefix{URL }\fi
\providecommand{\bibinfo}[2]{#2}
\providecommand{\eprint}[2][]{\url{#2}}

\bibitem[{\citenamefont{Bardeen}(1980)}]{Bardeen:1980kt}
\bibinfo{author}{\bibfnamefont{J.~M.} \bibnamefont{Bardeen}},
  \bibinfo{journal}{Phys. Rev.} \textbf{\bibinfo{volume}{D22}},
  \bibinfo{pages}{1882} (\bibinfo{year}{1980}).

\bibitem[{\citenamefont{Kodama and Sasaki}(1984)}]{Kodama:1985bj}
\bibinfo{author}{\bibfnamefont{H.}~\bibnamefont{Kodama}} \bibnamefont{and}
  \bibinfo{author}{\bibfnamefont{M.}~\bibnamefont{Sasaki}},
  \bibinfo{journal}{Prog. Theor. Phys. Suppl.} \textbf{\bibinfo{volume}{78}},
  \bibinfo{pages}{1} (\bibinfo{year}{1984}).

\bibitem[{\citenamefont{Baumgarte and Shapiro}(1999)}]{Baumgarte:1998te}
\bibinfo{author}{\bibfnamefont{T.~W.} \bibnamefont{Baumgarte}}
  \bibnamefont{and} \bibinfo{author}{\bibfnamefont{S.~L.}
  \bibnamefont{Shapiro}}, \bibinfo{journal}{Phys. Rev.}
  \textbf{\bibinfo{volume}{D59}}, \bibinfo{pages}{024007}
  (\bibinfo{year}{1999}), \eprint{gr-qc/9810065}.

\bibitem[{\citenamefont{Shibata and Nakamura}(1995)}]{Shibata:1995we}
\bibinfo{author}{\bibfnamefont{M.}~\bibnamefont{Shibata}} \bibnamefont{and}
  \bibinfo{author}{\bibfnamefont{T.}~\bibnamefont{Nakamura}},
  \bibinfo{journal}{Phys. Rev.} \textbf{\bibinfo{volume}{D52}},
  \bibinfo{pages}{5428} (\bibinfo{year}{1995}).

\bibitem[{\citenamefont{Clough et~al.}(2015)\citenamefont{Clough, Figueras,
  Finkel, Kunesch, Lim, and Tunyasuvunakool}}]{Clough:2015sqa}
\bibinfo{author}{\bibfnamefont{K.}~\bibnamefont{Clough}},
  \bibinfo{author}{\bibfnamefont{P.}~\bibnamefont{Figueras}},
  \bibinfo{author}{\bibfnamefont{H.}~\bibnamefont{Finkel}},
  \bibinfo{author}{\bibfnamefont{M.}~\bibnamefont{Kunesch}},
  \bibinfo{author}{\bibfnamefont{E.~A.} \bibnamefont{Lim}}, \bibnamefont{and}
  \bibinfo{author}{\bibfnamefont{S.}~\bibnamefont{Tunyasuvunakool}}
  (\bibinfo{year}{2015}), \eprint{1503.03436}.

\bibitem[{\citenamefont{Baumgarte and Montero}(2015)}]{Baumgarte:2015aza}
\bibinfo{author}{\bibfnamefont{T.~W.} \bibnamefont{Baumgarte}}
  \bibnamefont{and} \bibinfo{author}{\bibfnamefont{P.~J.}
  \bibnamefont{Montero}} (\bibinfo{year}{2015}), \eprint{1509.08730}.

\bibitem[{\citenamefont{Bentivegna and Korzynski}(2012)}]{Bentivegna:2012ei}
\bibinfo{author}{\bibfnamefont{E.}~\bibnamefont{Bentivegna}} \bibnamefont{and}
  \bibinfo{author}{\bibfnamefont{M.}~\bibnamefont{Korzynski}},
  \bibinfo{journal}{Class. Quant. Grav.} \textbf{\bibinfo{volume}{29}},
  \bibinfo{pages}{165007} (\bibinfo{year}{2012}), \eprint{1204.3568}.

\bibitem[{\citenamefont{Cardoso et~al.}(2015)\citenamefont{Cardoso, Gualtieri,
  Herdeiro, and Sperhake}}]{Cardoso:2014uka}
\bibinfo{author}{\bibfnamefont{V.}~\bibnamefont{Cardoso}},
  \bibinfo{author}{\bibfnamefont{L.}~\bibnamefont{Gualtieri}},
  \bibinfo{author}{\bibfnamefont{C.}~\bibnamefont{Herdeiro}}, \bibnamefont{and}
  \bibinfo{author}{\bibfnamefont{U.}~\bibnamefont{Sperhake}},
  \bibinfo{journal}{Living Rev. Relativity} \textbf{\bibinfo{volume}{18}},
  \bibinfo{pages}{1} (\bibinfo{year}{2015}), \eprint{1409.0014}.

\bibitem[{\citenamefont{Arnowitt et~al.}(1959)\citenamefont{Arnowitt, Deser,
  and Misner}}]{Arnowitt:1959ah}
\bibinfo{author}{\bibfnamefont{R.~L.} \bibnamefont{Arnowitt}},
  \bibinfo{author}{\bibfnamefont{S.}~\bibnamefont{Deser}}, \bibnamefont{and}
  \bibinfo{author}{\bibfnamefont{C.~W.} \bibnamefont{Misner}},
  \bibinfo{journal}{Phys. Rev.} \textbf{\bibinfo{volume}{116}},
  \bibinfo{pages}{1322} (\bibinfo{year}{1959}).

\bibitem[{\citenamefont{Baumgarte and Shapiro}(2010)}]{BaumgarteShapiroBook}
\bibinfo{author}{\bibfnamefont{T.~W.} \bibnamefont{Baumgarte}}
  \bibnamefont{and} \bibinfo{author}{\bibfnamefont{S.~L.}
  \bibnamefont{Shapiro}}, \emph{\bibinfo{title}{Numerical Relativity: Solving
  Einstein's Equations on the Computer}} (\bibinfo{publisher}{Cambridge
  University Press}, \bibinfo{address}{Cambridge, UK}, \bibinfo{year}{2010}).

\bibitem[{\citenamefont{Font}(2000)}]{Font:2000pp}
\bibinfo{author}{\bibfnamefont{J.~A.} \bibnamefont{Font}},
  \bibinfo{journal}{Living Rev. Rel.} \textbf{\bibinfo{volume}{3}},
  \bibinfo{pages}{2} (\bibinfo{year}{2000}), \bibinfo{note}{[Living Rev.
  Rel.6,4(2003)]}, \eprint{gr-qc/0003101}.

\bibitem[{\citenamefont{Duez et~al.}(2004)\citenamefont{Duez, Shapiro, and
  Yo}}]{Duez:2004uh}
\bibinfo{author}{\bibfnamefont{M.~D.} \bibnamefont{Duez}},
  \bibinfo{author}{\bibfnamefont{S.~L.} \bibnamefont{Shapiro}},
  \bibnamefont{and} \bibinfo{author}{\bibfnamefont{H.-J.} \bibnamefont{Yo}},
  \bibinfo{journal}{Phys. Rev.} \textbf{\bibinfo{volume}{D69}},
  \bibinfo{pages}{104016} (\bibinfo{year}{2004}), \eprint{gr-qc/0401076}.

\bibitem[{\citenamefont{Tegmark et~al.}(2004)}]{Tegmark:2003ud}
\bibinfo{author}{\bibfnamefont{M.}~\bibnamefont{Tegmark}} \bibnamefont{et~al.}
  (\bibinfo{collaboration}{SDSS}), \bibinfo{journal}{Phys. Rev.}
  \textbf{\bibinfo{volume}{D69}}, \bibinfo{pages}{103501}
  (\bibinfo{year}{2004}), \eprint{astro-ph/0310723}.

\bibitem[{\citenamefont{Dagum and Menon}(1998)}]{dagum1998openmp}
\bibinfo{author}{\bibfnamefont{L.}~\bibnamefont{Dagum}} \bibnamefont{and}
  \bibinfo{author}{\bibfnamefont{R.}~\bibnamefont{Menon}},
  \bibinfo{journal}{Computational Science \& Engineering, IEEE}
  \textbf{\bibinfo{volume}{5}}, \bibinfo{pages}{46} (\bibinfo{year}{1998}).

\bibitem[{\citenamefont{Frigo and Johnson}(2005)}]{FFTW05}
\bibinfo{author}{\bibfnamefont{M.}~\bibnamefont{Frigo}} \bibnamefont{and}
  \bibinfo{author}{\bibfnamefont{S.~G.} \bibnamefont{Johnson}},
  \bibinfo{journal}{Proceedings of the IEEE} \textbf{\bibinfo{volume}{93}},
  \bibinfo{pages}{216} (\bibinfo{year}{2005}), \bibinfo{note}{special issue on
  ``Program Generation, Optimization, and Platform Adaptation''}.

\bibitem[{\citenamefont{{The HDF Group}}(1997-NNNN)}]{hdf5}
\bibinfo{author}{\bibnamefont{{The HDF Group}}},
  \emph{\bibinfo{title}{{Hierarchical Data Format, version 5}}}
  (\bibinfo{year}{1997-NNNN}), \bibinfo{note}{http://www.hdfgroup.org/HDF5/}.

\bibitem[{\citenamefont{College}(2015)}]{tomweb}
\bibinfo{author}{\bibfnamefont{K.}~\bibnamefont{College}},
  \emph{\bibinfo{title}{Cosmology and computing at kenyon college}}
  (\bibinfo{year}{2015}), \urlprefix\url{http://cosmo.kenyon.edu}.

\bibitem[{EinsteinToolkit()}]{EinsteinToolkit:web}
EinsteinToolkit, \emph{\bibinfo{title}{{Einstein Toolkit}: Open software for
  relativistic astrophysics}}, \urlprefix\url{http://einsteintoolkit.org/}.

\bibitem[{\citenamefont{L{\"{o}}ffler et~al.}(2012)\citenamefont{L{\"{o}}ffler,
  Faber, Bentivegna, Bode, Diener, Haas, Hinder, Mundim, Ott, Schnetter
  et~al.}}]{Loffler:2011ay}
\bibinfo{author}{\bibfnamefont{F.}~\bibnamefont{L{\"{o}}ffler}},
  \bibinfo{author}{\bibfnamefont{J.}~\bibnamefont{Faber}},
  \bibinfo{author}{\bibfnamefont{E.}~\bibnamefont{Bentivegna}},
  \bibinfo{author}{\bibfnamefont{T.}~\bibnamefont{Bode}},
  \bibinfo{author}{\bibfnamefont{P.}~\bibnamefont{Diener}},
  \bibinfo{author}{\bibfnamefont{R.}~\bibnamefont{Haas}},
  \bibinfo{author}{\bibfnamefont{I.}~\bibnamefont{Hinder}},
  \bibinfo{author}{\bibfnamefont{B.~C.} \bibnamefont{Mundim}},
  \bibinfo{author}{\bibfnamefont{C.~D.} \bibnamefont{Ott}},
  \bibinfo{author}{\bibfnamefont{E.}~\bibnamefont{Schnetter}},
  \bibnamefont{et~al.}, \bibinfo{journal}{Class. Quantum Grav.}
  \textbf{\bibinfo{volume}{29}}, \bibinfo{pages}{115001}
  (\bibinfo{year}{2012}), \eprint{arXiv:1111.3344 [gr-qc]}.

\bibitem[{\citenamefont{Babiuc et~al.}(2008)}]{Babiuc:2007vr}
\bibinfo{author}{\bibfnamefont{M.~C.} \bibnamefont{Babiuc}}
  \bibnamefont{et~al.}, \bibinfo{journal}{Class. Quant. Grav.}
  \textbf{\bibinfo{volume}{25}}, \bibinfo{pages}{125012}
  (\bibinfo{year}{2008}), \eprint{0709.3559}.

\bibitem[{\citenamefont{Baumgarte and Naculich}(2007)}]{Baumgarte:2007ht}
\bibinfo{author}{\bibfnamefont{T.~W.} \bibnamefont{Baumgarte}}
  \bibnamefont{and} \bibinfo{author}{\bibfnamefont{S.~G.}
  \bibnamefont{Naculich}}, \bibinfo{journal}{Phys. Rev.}
  \textbf{\bibinfo{volume}{D75}}, \bibinfo{pages}{067502}
  (\bibinfo{year}{2007}), \eprint{gr-qc/0701037}.

\bibitem[{\citenamefont{Campanelli et~al.}(2006)\citenamefont{Campanelli,
  Lousto, Marronetti, and Zlochower}}]{Campanelli:2005dd}
\bibinfo{author}{\bibfnamefont{M.}~\bibnamefont{Campanelli}},
  \bibinfo{author}{\bibfnamefont{C.~O.} \bibnamefont{Lousto}},
  \bibinfo{author}{\bibfnamefont{P.}~\bibnamefont{Marronetti}},
  \bibnamefont{and}
  \bibinfo{author}{\bibfnamefont{Y.}~\bibnamefont{Zlochower}},
  \bibinfo{journal}{Phys. Rev. Lett.} \textbf{\bibinfo{volume}{96}},
  \bibinfo{pages}{111101} (\bibinfo{year}{2006}), \eprint{gr-qc/0511048}.

\bibitem[{\citenamefont{Ade et~al.}(2015)}]{Ade:2015xua}
\bibinfo{author}{\bibfnamefont{P.~A.~R.} \bibnamefont{Ade}}
  \bibnamefont{et~al.} (\bibinfo{collaboration}{Planck})
  (\bibinfo{year}{2015}), \eprint{1502.01589}.

\bibitem[{\citenamefont{Giblin et~al.}(2015)\citenamefont{Giblin, Mertens, and
  Starkman}}]{Giblin:2015vwq}
\bibinfo{author}{\bibfnamefont{J.~T.} \bibnamefont{Giblin}},
  \bibinfo{author}{\bibfnamefont{J.~B.} \bibnamefont{Mertens}},
  \bibnamefont{and} \bibinfo{author}{\bibfnamefont{G.~D.}
  \bibnamefont{Starkman}} (\bibinfo{year}{2015}), \eprint{1511.01105}.

\end{thebibliography}
  
\end{document}